\documentclass[final]{elsarticle}

\usepackage{hyperref}

\journal{International Journal of Thermal Sciences}

\usepackage[utf8]{inputenc}
\usepackage[english]{babel}
\usepackage{graphicx}
\usepackage{blindtext}

\usepackage[T1]{fontenc}
\usepackage{mathptmx}

\usepackage{hyperref}
\usepackage{amsmath}
\usepackage{xfrac}
\usepackage[inline]{enumitem}   
\usepackage{gensymb}
\usepackage{amssymb}
\usepackage{rotating}
\usepackage{relsize}
\usepackage{cleveref}
\usepackage{graphicx}
\usepackage{multirow}
\usepackage{booktabs}

\immediate\write18{%
	makeindex -s nomencl.ist -o \jobname.nls -t   \jobname.nlg \jobname.nlo%
}

\newcommand*\Fo{\mathrm{Fo}}

\usepackage{subfiles} 

\usepackage{nomencl}
\usepackage{multicol}
\makenomenclature

\usepackage{etoolbox}
\renewcommand\nomgroup[1]{%
	\item[\bfseries
	\ifstrequal{#1}{A}{Constants}{%
		\ifstrequal{#1}{B}{Heat equation}{%
			\ifstrequal{#1}{C}{Radiative transfer}{%
				\ifstrequal{#1}{D}{Discrete methods}{
					\ifstrequal{#1}{E}{Subscripts}{
						\ifstrequal{#1}{F}{Superscripts}{}}}}}%
	}]}

\begin{document}

\begin{frontmatter}

\title{Complexity matters: highly-accurate numerical models of coupled radiative-conductive heat transfer in a laser flash experiment}

\author[ukaea]{Artem Lunev\corref{mycorrespondingauthor}}
\cortext[mycorrespondingauthor]{Corresponding author}
\ead{artem.lunev@ukaea.uk}

\author[fian]{Vadim Zborovskii}
\author[fian]{Teymur Aliev}

\address[ukaea]{United Kingdom Atomic Energy Authority~(UKAEA), Culham Science Centre, Abingdon, Oxfordshire OX14 3DB, United Kingdom}
\address[fian]{P.N. Lebedev Physical Institute, Russian Academy of Sciences, Leninsky Prospect 53, Moscow 119991, Russia}

\begin{abstract}
Thermal diffusivity measurements of samples transmitting thermal radiation require adjustments to the data treatment procedures in laser flash analysis. Conventionally, an unconstrained diathermic model is used. Current results show that the alternative coupled radiative-conductive models produce substantially different results -- for instance, at high temperatures in oxide ceramics. However, care must be taken to ensure accurate implementations of each constituent computational technique. The latter are presented in this work. 
\end{abstract}

\begin{keyword}
	radiative transfer, heat conduction, discrete ordinates method, thermal diffusivity
\end{keyword}

\end{frontmatter}
\mbox{}


\section{Introduction}
\label{sect:intro}

High-temperature measurements of thermal properties using different experimental techniques such as the laser flash analysis~\cite{Pavlov2017} and the guarded hot plate method~\cite{Zhao2019} can be challenging for many reasons -- including, for instance, the stability of data acquisition and detector performance~\cite{Lunev2020}. When conducting tests on materials transmitting thermal radiation~(hereinafter referred to as the semi-transparent materials), e.g. metal oxides~\cite{Olo2002,Ita2006}, the optical properties of the sample material can significantly influence the measurement accuracy. Interestingly, even oxide nuclear fuel exhibits a degree of transparency to thermal radiation at high temperatures, potentially influencing the measurement procedure~\cite{Cozzo2011}. Other applications include thermal barrier coatings for the aerospace industry~\cite{Bison2007}. The non-vanishing interest in accurately measuring thermal properties of semi-transparent materials has instigated the development of mathematical methods aimed at quantifying radiative transfer and its coupling with heat conduction. Although some authors focussed on delivering quick estimates based on non-coupled heat conduction and radiative transfer~\cite{Tischler1988,McMasters1999,blumm1997laser,Mehling1998}, significant effort has been undertaken to address the coupled problem, primarily based on the works~\citep{Andre1995,Andre1998,Lazard2001,Lazard2001_2} with a recent development reported by~\citet{Braiek2016}. These models concern radiative transfer in either non-scattering or weakly-scattering media. An approximate solution to the radiative transfer equation~(RTE) using the exponential kernel technique and the two-flux method has been given for the latter case. These two methods had been used extensively in the past as follows from the introduction to~\cite{Modest1980} and could have hardly yielded realistic results for a scattering phase function with strong anisotropy. Moreover, the heating term was deemed small compared to the ambient temperature, which facilitated the solution of the initial problem. This is typically never satisfied under experimental conditions. To treat the radiative part, the three-flux method has also been considered in~\cite{Hahn1997} and an early attempt to use the discrete ordinates method~(DOM) -- first introduced by~\citet{Chandrasekhar1960} -- was reported by~\citet{Silva1998}. Currently, DOM is often applied to this kind of problems~\cite{Coquard2009,Coquard2011,Wel2006}, although it is still not clear if this has indeed increased the reliability of laser flash analysis on semi-transparent samples. Even using the DOM formalism, some authors still defer to non-scattering transfer blaming difficulties in the estimation of some coefficients~\cite{Sans2020}. \citet{zmywaczyk2009numerical,Lacrois2002} have used the DOM in its rather conventional form with reference to~\citet{Fiveland1984,Fiveland1987}. On the other hand, simplified non-coupled models are still the most popular choice for experimental data treatment in the majority of cases~\cite{philipp2020accuracy}. It is thus inconclusive if solving a coupled conductive-radiative problem is advantageous compared to using less demanding methods, and whether anisotropic scattering has any measurable effect on the thermal properties determined from a laser flash experiment. This is partially due to numerical heat transfer still being a developing area with many caveats still not addressed sufficiently: particularly, for the spatial and angular discretisation in DOM~\cite{Coelho2007,Coelho2014}. This paper is aimed at delivering reliable and fast numerical algorithms for one-dimensional coupled conductive-radiative heat transfer with application to the laser flash analysis. The algorithm and procedures outlined in this work are part of the PULsE~(\textbf{P}rocessing \textbf{U}nit for \textbf{L}aser Fla\textbf{s}h \textbf{E}xperiments) software, which is an open-source, cross-platform Java code freely distributed under the Apache 2.0 license~\cite{software}.

\mbox{}
\nomenclature[A, 01]{$\sigma_0$}{Stefan–Boltzmann constant}
\nomenclature[B, 14]{$y=z/l$}{Dimensionless coordinate}
\nomenclature[B, 01]{$l$}{Sample thickness}
\nomenclature[B, 13]{$F$}{Heat flux}
\nomenclature[B, 02]{$T_0$}{Ambient temperature}
\nomenclature[B, 10]{$T(z,t)$}{Local temperature}
\nomenclature[B, 09]{$\theta = (T - T_0)/{\delta T_m}$}{Dimensionless heating}
\nomenclature[B, 08]{$\delta T_m = 4Q/(C_{\mathrm{p}} \rho \pi d^2 l)$}{Adiabatic heating}
\nomenclature[B, 03]{$Q$}{Energy per laser pulse}
\nomenclature[B, 04]{$d$}{Sample diameter}
\nomenclature[B, 05]{$\varepsilon$}{Hemispherical emissivity}
\nomenclature[B, 15]{$\eta = \varepsilon/(2 - \varepsilon)$}{Diathermic coefficient}
\nomenclature[B, 12]{$a$}{Thermal diffusivity}
\nomenclature[B, 06]{$\lambda$}{Thermal conductivity}
\nomenclature[B, 07]{$\mathrm{Bi} = 4 \sigma_0 \varepsilon T_0^3 l/\lambda$}{Biot number}
\nomenclature[B, 11]{$\mathrm{Fo}$}{Fourier number}
\nomenclature[C, 06]{$q = {F}/{(n^2\sigma_0 T_0^3)}$}{Dimensionless heat flux}
\nomenclature[C, 09]{$I$, $i$}{Intensity (dimensionless)}
\nomenclature[C, 07]{$\tau_0 = l \psi$}{Optical thickness}
\nomenclature[C, 08]{$\tau = \tau_0 y$}{Optical coordinate}
\nomenclature[C, 01]{$\psi$}{Hemispherical absorptivity}
\nomenclature[C, 02]{$\chi$}{Hemispherical scattering coefficient}
\nomenclature[C, 11]{$\Phi(\mu,\mu')$}{Scattering phase function}
\nomenclature[C, 10]{$\mu$, $\mu'$}{Direction cosine of incident (scattered) rays}
\nomenclature[C, 15]{$S$, $s$}{Source function (dimensionless)}
\nomenclature[C, 03]{$\omega_0$}{Single-scattering albedo}
\nomenclature[C, 13]{$N_{\mathrm{P}} = \lambda/(4 \sigma_0 n^2 T_0^3 l)$}{Planck number}
\nomenclature[C, 04]{$n$}{Refractive index}
\nomenclature[C, 12]{$J(t)$, $j(t)$}{Integrated spectral radiance (dimensionless)}
\nomenclature[C, 14]{$E_n(t)$}{Exponential integral of the order~$n$}
\nomenclature[C, 05]{$g$}{Scattering anisotropy}
\nomenclature[D]{$w$}{Quadrature weights}
\nomenclature[D]{$M$}{Number of quadrature nodes}
\nomenclature[D]{$\Delta t$}{$\mathrm{Fo}$ increment (time step)}
\nomenclature[D]{$N$}{Number of spatial grid points}
\nomenclature[D]{$\sigma$}{Scheme weight}
\nomenclature[D]{$f$}{Right-hand side (RTE)}
\nomenclature[D]{$a_{nn'}$, $b_n$, $\widehat{b}_n$, $c_n$}{Butcher tableau coefficients}
\nomenclature[D]{$\xi_j$}{Grid point}
\nomenclature[D]{$h$}{Uniform grid step}
\nomenclature[D]{$h_l$}{Adaptive grid step}
\nomenclature[D]{$t$}{Discrete optical coordinate}
\nomenclature[D]{$\mathbf{Est}$, $\mathbf{est}$}{Error estimators}
\nomenclature[D]{$\phi$}{Discrete flux derivative}
\nomenclature[D]{$atol$}{Absolute error tolerance}
\nomenclature[D]{$rtol$}{Relative error tolerance}
\nomenclature[D]{$s_G$}{Stretching factor}
\nomenclature[D]{$\tau_{F}$}{Time step factor}
\nomenclature[D]{$\omega_R$}{Relaxation parameter}
\nomenclature[D]{$e_{\mathrm{it}}$}{Relaxation error tolerance of iterative solution}
\nomenclature[D]{$L$}{Central-difference operator}
\nomenclature[D]{$\Lambda$}{Second-order difference operator}
\nomenclature[E]{$m$, $m'$}{Angular indices (DOM)}
\nomenclature[E]{$l$}{Spatial index (DOM)}
\nomenclature[E]{$j$}{Spatial index (heat equation)}
\nomenclature[F]{$n$, $s$}{Stage number}
\nomenclature[F]{$i$}{Time step number}
\nomenclature[F]{$\widehat{}$}{Value at previous time step}
\nomenclature[F]{$k$, $u$}{Iteration numbers}

\begin{multicols}{2}
\footnotesize{
\printnomenclature
}
\end{multicols}

\section{Diathermic medium bounded by grey walls}
\label{sect:diathermic}

\subsection{Problem statement}
\label{sect:diathermic_statement}

Early models used in laser flash measurements of semi-transparent samples considered radiation and conduction as non-coupled phenomena since this greatly simplifies the mathematical formulation of the problem. \citet{Tischler1988} considered an exponential decay of radiation intensity in a solid partially transparent to the laser pulse. \citet{McMasters1999} applied the optically thick approximation and introduced an additional source term in the heat equation. These models are useful to gain a crude estimate of thermal diffusivity e.g. in porous samples and semi-conductors with an intermediate band gap. Rather than considering laser penetration in solids -- a complex problem associated with the diffusion of charge carriers, their re-combination and thermalisation by phonon emission~\cite{Yoffa1980} -- it is much easier to manually restrict the laser absorption depth by applying a graphite coating. ~\citet{blumm1997laser} proposed the diathermic model specifically to deal with this case; an analytical solution was later developed by~\citet{Mehling1998}. A variation of this model is currently being used in software packaged with some commercial instruments. 

The diathermic model is based on the following propositions:

\begin{enumerate}[label=(\alph*)]
	\item A cylindrically shaped sample is completely transparent to thermal radiation;
	\item The front~(laser-facing) and rear~(detector-facing) sides of the sample are coated by a thin grey absorber;
	\item The coatings are in perfect thermal contact with the bulk material;
	\item The side surface is free from any coating.
\end{enumerate} 

Consequently, the monochromatic laser radiation is largely absorbed at the front face of the sample~($y = 0$), causing immediate heating. A portion of thermal radiation causes the rear face~($y = 1$) to start heating precisely at the same time~(ahead of thermal conduction). The remainder energy dissipates in the ambient. It is thus sufficient to consider three radiative heat fluxes. The first two correspond to heat dissipation within the furnace chamber~\cite{Lunev2020}. The third flux acts to thermalise the parallel boundaries by radiative transfer only~\cite{howell2010thermal}:

\begin{subequations}
	\label{eq:radiation_fluxes}
	\begin{align}
	\label{eq:cooling_flux_front}
	&F_{0 \to \infty} \approx \varepsilon \sigma_0 \left ( T^4(0,t) - T_0^4 \right ),\\
	\label{eq:cooling_flux_rear}
	&F_{1 \to \infty} \approx - \varepsilon \sigma_0 \left ( T^4(l,t) - T_0^4 \right ),\\
	\label{eq:diathermic_flux_12}
	&F_{1 \to 2} \approx \frac{\varepsilon}{2 - \varepsilon} \sigma_0 \left ( T^4(0,t) - T^4(l,t) \right ),
	\end{align}	
\end{subequations}
where the emissivities of both faces are assumed to be equal~($\varepsilon_1 = \varepsilon_2 = \varepsilon$). 

Let~$\eta = \varepsilon/(2 - \varepsilon)$, so that $0 < \eta \leq 1$. Since nonlinear heat losses can be neglected~[\ref{sect:appendix_a}], the boundary problem is written as:

\begin{subequations}
	\label{eq:boundary_problem_diathermic}
	\begin{gather}
	\label{eq:he_nodim}
	\frac{{\partial \theta }}{{\partial {\rm{Fo}}}} = \frac{{{\partial ^2}\theta }}{{\partial {y^2}}},\quad {\kern 1pt} 0 < y < 1,\quad {\rm{Fo}} > 0,\\
	\label{eq:bc_0}
	{\left. {\frac{{\partial \theta }}{{\partial y}}} \right|_{y = 0}} = {\rm{Bi}} \cdot \theta_{y = 0} + \eta {\rm{Bi}} \cdot (\theta_{y = 0} - \theta_{y = 1}) - \Phi \left( \rm{Fo} \right),\\
	\label{eq:bc_1}
	{\left. {\frac{{\partial \theta }}{{\partial (-y)}}} \right|_{y = 1}} = {\rm{Bi}} \cdot \theta_{y = 1} + \eta {\rm{Bi}} \cdot (\theta_{y = 1} - \theta_{y = 0}), \\
	\label{eq:ic_nodim}
	\theta(0,y) = 0,
	\end{gather}
\end{subequations}
where~\cref{eq:he_nodim,eq:ic_nodim} and the corresponding notations are the same as in~\cite{Lunev2020}. The standard non-dimensional variables are used, also defined in the same reference.

\subsection{A finite-difference solution}

Let the superscript~$i$ and the subscript~$j=0,...,N-1$ denote the time step and the coordinate index respectively. The boundary conditions~[\cref{eq:bc_0,eq:bc_1}] are expressed in finite differences as follows:

\begin{subequations}
	\label{eq:boundary_problem_diathermic_fdm}
	\begin{align}
	&L \theta_0 = \mathrm{Bi} \cdot \theta_0 + \eta \mathrm{Bi} \cdot (\theta_0 - \theta_{N-1}) - \widetilde{\Phi}^{i+1},\\
	&-L \theta_{N-1} = \mathrm{Bi} \cdot \theta_{N-1} + \eta \mathrm{Bi} \cdot (\theta_{N-1} - \theta_0),
	\end{align}	
\end{subequations}

The usual Taylor expansion is written down in the $h$-vicinity of~$\xi = \xi_0$ and $\xi = \xi_{N-1}$, thus defining the virtual nodes~$\xi = \xi_{-1}$ and $\xi = \xi_N$ needed to evaluate the boundary derivatives. After some elementary algebra, an $O(h^2 + \Delta t^2)$ accurate scheme is readily obtained:

\begin{subequations}
	\label{eq:bc_diathermic_fdm}
	\begin{align}
	&\theta_0^{i+1} \left[ 1 + h^2/(2 \Delta t) + h \mathrm{Bi} (1 + \eta) \right] 
	 - \theta_1^{i+1} -  \theta_N^{i+1} h \eta \mathrm{Bi} = h^2/(2 \Delta t) \theta_0^i + h \Phi^{i+1}, \\
	&\theta_{N-1}^{i+1} \left[ 1 + h^2/(2 \Delta t) + h\mathrm{Bi} (1 + \eta) \right] 
	 - \theta_{N-2}^{i+1} -  \theta_0^{i+1} h \eta \mathrm{Bi} = h^2/(2 \Delta t) \theta_{N-1}^i, 
	\end{align} 
\end{subequations}

with the heat equation also given in finite differences:

\begin{equation}
\label{eq:implicit_scheme}
a_j \theta_{j-1}^{i+1} - b_j \theta_j^{i+1} + c_j \theta_{j+1}^{i+1} = R_j,
\end{equation}
where $a_j = c_j = 1$. A fully implicit scheme shown previously to work well in most cases~\cite{Lunev2020} corresponds to: $b_j = 2 + h^2/{\Delta t}$, $R_j = - h^2 / {\Delta t} \theta_j^i$.

\Cref{eq:bc_diathermic_fdm,eq:implicit_scheme} are reduced to the following linear matrix equation:

\begin{subequations}
\label{eq:matrix_fdm_diathermic}
\begin{gather}
\label{eq:matrix_equation_1}
\mathbf{A} \underline{\theta}^{i+1} = \mathbf{R}, \\ 
\mathbf{A} = 
\begin{pmatrix}
z_0 & -1 & 0 & 0 & 0 & \cdots & 0 & 0 & z_{N-1}  \\
a_1 & -b_1 & c_1 & 0 & 0 & \cdots & 0 & 0 & 0 \\
0 & a_2 & -b_2 & c_2 & 0 & \cdots & 0 & 0 & 0 \\
\vdots & \vdots & \vdots & \vdots & \vdots & \ddots & \vdots & \vdots & \vdots \\
0 & 0 & 0 & 0 & 0 & \cdots & a_{N-2} & -b_{N-2} & c_{N-2} \\
z_{N-1} & 0 & 0 & 0 & 0 & \cdots & 0 & -1 & z_0 
\end{pmatrix}, \\
\mathbf{R}^{\mathbf{T}} = \begin{pmatrix} 
r_0 & R_1 & R_2 & \cdots & R_{N-2} & r_{N-1}
\end{pmatrix}
\end{gather}
\end{subequations}
where~$\underline{\theta}^{\mathbf{T}} = \begin{pmatrix} 
\theta_0 & \theta_1 & \cdots & \theta_{N-1} \end{pmatrix}$, $z_0 = 1 + h^2/(2 \Delta t) + h \mathrm{Bi} (1+\eta)$, $z_{N-1} = -h \eta \mathrm{Bi}$, $r_0 = h^2/(2\Delta t) \theta_0^i + h \Phi^{i+1}$ and~$r_{N-1} = h^2/(2\Delta t) \theta_{N-1}^i$. Since~$z_{N-1} \neq 0$, the matrix~$\mathbf{A}$ does not have the required tridiagonal form. This problem can be solved by applying the bordering method~\cite{samarskii1978methods}. Consider the following equations equivalent to Eq.~\eqref{eq:matrix_equation_1}:

\begin{subequations}
	\begin{align}
	\label{eq:first_border_matrix_eq}
	&\mathbf{A'} \underline{\underline{\theta}}^{i+1} + \mathbf{u'} \underline{\underline{\theta}}^{i+1} = \mathbf{R'}, \\ 
	&\mathbf{{v'}^T} \underline{\underline{\theta}}^{i+1} + z_0 \theta_{N-1} = r_{N-1},
	\end{align}
\end{subequations}

where~$\mathbf{A'} = \mathbf{A}_{N-1,N-1}$ is the border minor of the matrix~$\mathbf{A}$, the vectors $\underline{\underline{\theta}}^{\mathbf{T}} = \begin{pmatrix} 
\theta_0 & \theta_1 & \cdots & \theta_{N-2} \end{pmatrix}$ and $\mathbf{F'} = \begin{pmatrix}
r_0 & 0 & \cdots & 0 & R_{N-2}
\end{pmatrix}$. The vector
\begin{equation} 
\mathbf{u'^T} = 
\begin{pmatrix} z_{N-1} & 0 & \cdots & 0 & c_{N-2} 
\end{pmatrix} \nonumber
\end{equation}
is formed from the last column of~$\mathbf{A}$. Conversely,
\begin{equation}
\mathbf{{v'}^T} = \begin{pmatrix} 
z_{N-1} & 0 & \cdots & 0 & -1
\end{pmatrix} \nonumber
\end{equation} 
is formed from the last row of~${\mathbf{A}}$. 

A solution to Eq.~\eqref{eq:first_border_matrix_eq} is sought in the form~$\underline{\underline{\theta}}^{i+1} = \mathbf{w^I} + \theta_{N-1} \mathbf{w^{II}}$, where~$\mathbf{w^I}$ and~$\mathbf{w^{II}}$ are the solutions of the linear matrix equations~$\mathbf{A'} \mathbf{w^I} = \mathbf{R'}$ and~$\mathbf{A'} \mathbf{w^{II}} = -\mathbf{u'}$. Since~$\mathbf{A'}$ is a Jacobi matrix, the two latter equations can be solved using the standard tridiagonal matrix algorithm. The following relations hold:

\begin{equation}
w^\mathrm{I}_i = \alpha_{i+1} w^{\mathrm{I}}_{i+1} + \beta_{i+1}, \quad
w^{\mathrm{II}}_i = \alpha_{i+1} w^{\mathrm{II}}_{i+1} + \kappa_{i+1}.
\end{equation}

The sweep algorithm coefficients can then be easily calculated:

\begin{subequations}
\begin{gather}
\alpha_1 = 1/z_0, \quad \beta_1 = r_0/z_0, \quad \kappa_1 = -z_{N-1}/z_0 \\
\alpha_{j+1} = \frac{c_j}{b_j - \alpha_j a_j}, \quad \beta_{j+1} = \frac{R_j - a_j \beta_j}{a_j \alpha_j - b_j}, \quad \kappa_{j+1} = \frac{a_j \kappa_j}{b_j - a_j \alpha_j}, \quad j = 1,..., N-2 \\
w^{\mathrm{I}}_{N-2} = \beta_{N-1}, \quad w^{\mathrm{II}}_{N-2} = \alpha_{N-1} + \kappa_{N-1}, \\
w_j^{\mathrm{I}} = \alpha_{j+1}w^{\mathrm{I}}_{j+1} + \beta_{j+1}, \quad w_j^{\mathrm{II}} = \alpha_{j+1}w^{\mathrm{II}}_{j+1} + \kappa_{j+1}, \quad j = N - 3, ..., 0 \\
\theta_{N-1} = \frac{r_{N-1} - \mathbf{v'^T} \cdot \mathbf{w^I}}{ z_0 + \mathbf{v'^T} \cdot \mathbf{w^{II}} } = \frac{r_{N-1} - z_{N-1}w^{\mathrm{I}}_0 + w^{\mathrm{I}}_{N-2}}{z_0 + z_{N-1}w^{\mathrm{II}}_0 - w^{\mathrm{II}}_{N-2}}, \\ 
\theta_{j} = w^{\mathrm{I}}_j + \theta_{N-1} w^{\mathrm{II}}_j, \quad j = 0, ..., N-2. 
\end{gather}
\end{subequations}

\begin{figure}
	\centering
	\includegraphics{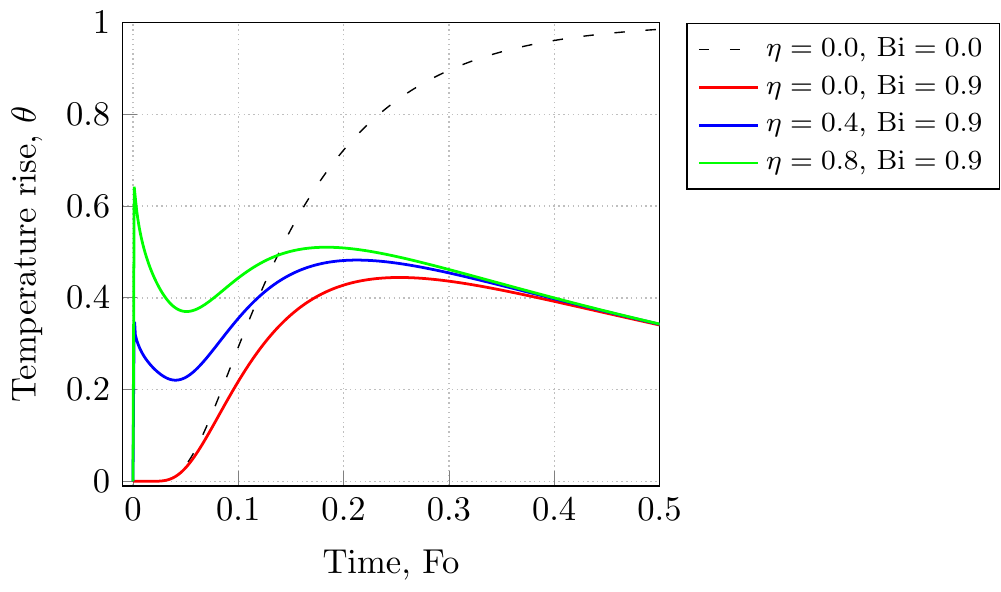}
	\caption{\label{fig:diathermic_example} An example calculation using the diathermic model~(\cref{sect:diathermic_statement}) at different~$\mathrm{Bi}$ and~$\eta$ values. The radiative terms in \cref{eq:bc_0,eq:bc_1} cause significant deviation from the classical behaviour.}
\end{figure}

An example calculation using the diathermic model with the finite-difference scheme described in this section is shown in~\cref{fig:diathermic_example}. The calculation used a default grid density~$N = 30$ and a time step~$\Delta t = t_{\mathrm{F}} h^2$, where~$t_{\mathrm{F}} = 0.5$.

\section{The general form of the coupled conductive-radiative heat transfer problem}
\label{sect:heat_problem}

The following is the equation of radiative transfer in a plane-parallel geometry with an axially symmetric radiation field for a grey participating~(i.e., emitting, absorbing, and scattering) medium compliant with the Kirchhoff's law~\cite{kourganoff1963basic}: 
	
		\begin{gather}
		\label{eq:radiative_transfer}
		\frac{d I}{d s} =  - (\psi +\chi) I + \psi J + \chi \int_{\mu'}{I \Phi (\mu,\mu') \frac{\mathrm{d}{\mu'}}{2}}, \\ 
		\label{eq:emission_function}
		J = J(\tau) = \frac{n^2 \sigma_0 T^4(\tau)}{\pi} = \frac{n^2 \sigma_0 T_0^4}{\pi} \left ( \frac{T(\tau) - T_0}{T_0} + 1 \right )^4,
		\end{gather}
where~$s$ is the path travelled by radiation; $n$ is the refractive index of the medium; $\psi$, $\chi$ and~$\varepsilon$ are respectively the linear absorption coefficient, the scattering coefficient and the emissivity -- all averaged over the radiation spectrum;~$\Phi(\mu,\mu')$ is the phase function of scattering, such that~$\int{\Phi(\mu,\mu')d\mu'}/2 = 1$; $\mu = \cos \Theta$ is the cosine of the angle between the light propagation direction and the outward normal to an elementary illuminated surface. 
	
It is convenient to express \cref{eq:radiative_transfer} in terms of the optical thickness~$\tau = \int{\psi \ \mathrm{d}s} = \int{\psi \ \mathrm{d}y/\cos \Theta}$, which then allows separating the positive~($0 < \mu \leq 1$) and negative~($-1 \leq \mu < 0$) streams. After introducing  the albedo for single scattering~$\omega_0 = \chi/(\psi + \chi)$, the RTE e.g. for $I^+$ takes the form: 
	
	\begin{equation}
	\label{eq:radiative_transfer_2}
	\mu \frac{\partial I^+}{\partial \tau} + I^+ = S(\tau,\mu), \quad 0 < \mu \leq 1, \\
	\end{equation}
where the source function is defined as~$S(\tau,\mu) = (1-\omega_0) J + 0.5 \omega_0 \int_{\mu'}{I \Phi (\mu,\mu') \mathrm{d}{\mu'}}$.
	
A matching equation may be written for~$I^-$, thus the RTE may bs solved separately for streams propagating in the positive and negative hemisphere originating at either~$\tau = 0$ or at $\tau = \tau_0 := l \psi$. The complexity of the problem is determined by the source function~$S(\tau,\mu)$, which in some cases, e.g. at~$\omega_0 = 0$, allows an analytical solution. Once a solution has been obtained, the net radiative heat flux~$F(\tau)$ can be calculated using an expression for a radiative field with axial symmetry~\cite{Chandrasekhar1960}:

\begin{equation}
\label{eq:radiative_flux}
F(\tau) = 2 \pi \int_{-1}^{1}{I(\mu, \tau) \mu d{\mu} } = 2 \pi \left [ \int_{0}^1{I^+(\mu) \mu \mathrm{d}{\mu}} - \int_{0}^{1}{I^-(-\mu) \mu \mathrm{d}{\mu}}  \right ],
\end{equation}

Conduction and radiation both contribute to the heat flow, which becomes

\begin{equation}
- \frac{\lambda}{l} \frac{\partial T}{\partial y} + F(\tau_0 y) \nonumber
\end{equation}

In the isotropic case~$dF/dy = \tau_0 \times dF/d\tau$ and the dimensionless heat equation may be written as: 

\begin{subequations}
\label{eq:coupled_problem}
\begin{gather}
\label{eq:coupled_problem_rte}
\frac{\partial \theta}{\partial \Fo} = \frac{\partial^2 \theta}{\partial y^2} + \frac{\tau_0}{N_{\mathrm{P}}}  \times \left (-\frac{dq}{d\tau} \right ), \\
\label{eq:bc_lower}
{\left. {\frac{{\partial \theta }}{{\partial y}}} \right|_{y = 0}} = {\rm{Bi}} \cdot \theta  - \Phi \left( \rm{Fo} \right) + \frac{1}{N_{\mathrm{P}}} q(0),\\
\label{eq:bc_lower_radiative_flux}
{\left. {\frac{{\partial \theta }}{{\partial y}}} \right|_{y = 1}} = -{\rm{Bi}} \cdot \theta + \frac{1}{N_{\mathrm{P}}} q(1),\\
\label{eq:bc_upper_radiative_flux}
\theta (0,y) = 0.
\end{gather}
\end{subequations}
where~$q = F /(n^2\sigma_0 T_0^3 )$ is the dimensionless radiative flux; in addition, the Planck number is introduced:~$N_\mathrm{P} = \lambda / \left (4 \sigma_0 n^2 T_0^3 l \right )$. The re-normalisation of the heat flux simply leads to substituting the emission function~$J(\tau)$ ~[\cref{eq:emission_function}] with

\begin{equation}
j(\tau) = \frac{1}{4\pi} \frac{T_0}{\delta T_{\mathrm{m}}} \left [ 1 + \theta \left ( \tau/\tau_0 \right ) \frac{\delta T_m}{T_0} \right ]^4,
\end{equation}
which is also dimensionless. 

The boundary radiative fluxes~$q(0)$ and~$q(1)$ are inferred from the boundary intensities~$I^+(0)$ and~$I^-(\tau_0)$ determined through the conditions of diffuse emission and reflection~[e.g. \cite{modest2013radiative}]:

\begin{subequations}
	\label{eq:rte_bc}
	\begin{align}
	I^+(0) &= \varepsilon J(0) + (1 - \varepsilon) G_0/\pi, \\
	I^-(\tau_0) &= \varepsilon J(\tau_0) + (1 - \varepsilon) G_{\tau_0}/\pi,
	\end{align}
\end{subequations}
where~$G_0$ and~$G_{\tau_0}$ is the incident irradiation reaching the respective boundary.

\section{A closer look at the radiation problem}

\subsection{Useful special cases}

\subsubsection{Exact solution at~$\omega_0 = 0$} 
\label{sect:rte_exact}	
	
In the absence of scattering, the source function~$s(\tau,\mu)$ is simply equal to the emission function~$j(\tau)$. This then simplifies the equation, which is solved in terms of the exponential integrals~$E_n(t)$~[see e.g. \cite{kourganoff1963basic}]. The latter are defined as~$E_n (t) = \int_0^1{e^{-t/\mu}\mu^{n-2}\mathrm{d}\mu}$, $n > 0$. This leads to the following expression for the radiative flux~\cite{Cess1964}:

\begin{subequations}
\label{eq:radiation_flux}
\begin{align}
\nonumber &q(\tau)/2= \pi i^+(0) E_3(\tau) - \pi i^-(\tau_0) E_3(\tau_0 - \tau) + \\
&\int_0^{\tau}{\pi j(t) E_2(\tau - t) dt} - \int_{\tau}^{\tau_0}{\pi j(t) E_2(t-\tau)dt},
\end{align}
\end{subequations}
where~$i^+(0)$ and~$i^-(0)$ are the boundary intensities.

Consequently, the radiation fluxes at the boundaries are:

\begin{subequations}
\label{eq:boundary_fluxes}
\begin{align}
q(0) = &\pi i^+(0) -2 \pi i^-(\tau_0) E_3(\tau_0) - 2 \int_{0}^{\tau_0}{\pi j(t) E_2(t) \mathrm{d}t }, \\ 
q(\tau_0) = &-\pi i^-(\tau_0) + 2 \pi i^+(0) E_3(\tau_0) + 2 \int_0^{\tau_0}{\pi j(t) E_2(\tau_0 - t) \mathrm{d}t}.
\end{align}
\end{subequations}

Combining \cref{eq:rte_bc,eq:boundary_fluxes} allows to evaluate~$i^+(0)$ and~$i^-(\tau_0)$ from a set of two linear equations~(see e.g. \cite{Andre1995}):
\begin{subequations}
	\label{eq:radiosities}
	\begin{align}
	&i^+(0) = \frac{C_1 + D C_2}{1 - D^2}, \quad i^-(\tau_0) = \frac{C_2 + D C_1}{1 - D^2}, \\
	&C_1 = \varepsilon j(0) + 2 (1 - \varepsilon) \int_0^{\tau_0}{ j(t) E_2(t)\mathrm{d}t}, \\
	&C_2 = \varepsilon j(\tau_0) + 2 (1 - \varepsilon) \int_0^{\tau_0}{ j(t) E_2 \left (\tau_0 - t \right )\mathrm{d}t}, \\
	&D = 2(1 -\varepsilon)E_3(\tau_0).
	\end{align}
\end{subequations}

The heat source term in~\cref{eq:coupled_problem_rte} can either be calculated using a discrete approximation or exactly using the analytic expression below derived using the properties of the exponential integral~(the reader is referred to~\cite{kourganoff1963basic}):
\begin{align}
\label{eq:analytic_deivatives}
&\left (-\frac{dq}{d\tau} \right ) = 2 i^+(0) E_2(\tau) + 2 i^-(\tau_0) E_2 \left ( \tau_0 - \tau \right ) \nonumber \\ &- 4 j(\tau) + 2  \int_0^{\tau_0}{j(t) E_1( |\tau - t|) \mathrm{d}t}
\end{align}

A quadrature scheme needs to be used in order to calculate the integrals in~\cref{eq:radiation_flux,eq:analytic_deivatives} -- this is given in~\ref{sect:appendix_b}.

\subsubsection{The two-flux approximation}
\label{sect:two-flux}

For a weakly-anisotropic phase function~$\Phi(\mu,\mu')$, when~$\tau_0$ is not very large, the two-flux approximation originally introduced by~\citet{schuster1905radiation,schwartzschild1906gottingen} has been shown to yield sufficiently accurate results~\cite{Brewster1982}. In current notations, this approximation considers~$I^+$ and~$I^-$ averaged over the positive and negative hemispheres correspondingly. The governing equations are then~\citep{Menguc1983}:

\begin{subequations}
\label{eq:two-flux}
\begin{align}
&\frac{dI^+}{d\tau} = -2I^+ [1 - (1 - u) \omega_0 ] + 2 \omega_0 u I^- + 2(1 - \omega_0) \pi J(\tau), \\
&\frac{dI^-}{d\tau} = -2I^- [1 - (1 - u) \omega_0 ] + 2 \omega_0 u I^+ + 2(1 - \omega_0) \pi J(\tau),
\end{align} 
\end{subequations}
where $u$ is an integral scattering parameter of the model.

The phase function can be expanded in a series of Legendre polynomials~\cite{Chandrasekhar1960}. In the linear-anisotropic approximation the series is truncated after the second term, which in a axially-symmetric radiation field gives rise to~$\Phi(\mu,\mu') = 1 + g \mu \mu'$. This corresponds to~$u = 0.5 - 0.25 g$.

\subsection{The general case of strong anisotropic scattering in a nonlinear grey participating medium}
\label{sect:rte_general}

The true multi-modal~\cite{Engler2015} form of the scattering function~$\Phi(\mu,\mu')$ can be derived from the Lorenz-Mie theory~(see e.g.~\cite{hulst1981light}). In most practical applications, it is more convenient to use an approximation, which still captures the strongly anisotropic scattering behaviour. This is commonly done using the single-parameter Henyey-Greenstein phase function~\cite{henyey1941diffuse}. Other specialised functions have been discussed in~\cite{kattawar1975three,Haltrin,Wang2019,zhao2019influence}. 

The phase function of interest is thus:

\begin{equation}
\label{eq:henyey_greenstein}
\Phi(\mu,\mu') = (1 - g^2)(1 + g^2  - 2 g \mu \mu')^{-3/2}.
\end{equation}   

If the integral over~$\Phi(\mu,\mu')$ in~\cref{eq:radiative_transfer} cannot be simplified, as in case of~$\Phi(\mu, \mu')$ given by~\cref{eq:henyey_greenstein}, the solution to the RTE becomes quite involved. Some effort in solving the RTE and, indeed, the coupled problem has been undertaken by many authors~\cite{Lacrois2002,Boulet2007,zmywaczyk2009numerical}. Generally, the discrete ordinates method~(DOM) is used for this purpose. Henceforth, the paper is focussed on the numerical implementation of DOM. 

Recall the general form of the source function:
\begin{align}
\label{eq:source_scattering}
s(\tau, \mu) = &(1 - \omega_0) j(\tau) + \frac{\omega_0}{2} \int_{-1}^1{i(\tau,\mu) \Phi(\mu,\mu') d\mu}
\end{align}

The idea behind DOM is to evaluate the integral on the right-hand side using a quadrature rule. A discrete set of nodes is introduced: $\mu_m$, $m = 0, ..., M - 1$, with an equal number of negative and positive nodes; each node is assigned a certain weight~$w_m$. The discrete form of \cref{eq:source_scattering} is:
\begin{align}
\label{eq:dom_source}
&s_m = (1 - \omega_0) j(\tau) + \frac{\omega_0}{2} \sum_{m' = 0}^{M - 1} {i_{m'}  \Phi(\mu_m,\mu_{m'}) w_{m'} \mu_{m'}}.
\end{align} 

The discrete RTE~[\cref{eq:radiative_transfer_2}] is given by:    
\begin{equation}
\label{eq:rte_dom}
\mu_m \frac{\partial i_m}{\partial \tau} + i_m = s_m
\end{equation}

with the boundary conditions of diffuse emission and reflection~[\cref{eq:rte_bc}]:

\begin{subequations}
\label{eq:bc_dom}
\begin{align}
i_m(0,\mu_m > 0) = &\varepsilon j(0) - 2 (1 - \varepsilon) \sum_{\mu_{m'} < 0} {i_{m'} \mu_{m'} w_{m'}} \\
i_m(\tau_0,\mu_m < 0) = &\varepsilon j(\tau_0) + 2 (1 - \varepsilon) \sum_{\mu_{m'} > 0} {i_{m'} \mu_{m'} w_{m'}}
\end{align}
\end{subequations}

The net radiative flux~[\cref{eq:radiation_flux}]:
\begin{equation}
\label{rq:flux_dom}
q(\tau) = 2 \pi \sum_{m= 0}^{M-1}{i_m \mu_m w_m}.
\end{equation}

\section{The solution to the heat problem}
\label{sect:schemes}

It is convenient to first select an appropriate numerical scheme for solving the heat problem outlined in~\cref{sect:heat_problem} before launching a full-scale analysis of the radiative transfer problem~(\cref{sect:rte_general}). For this reason, the analytical solution obtained in~\cref{sect:rte_exact} is used to calculate the heat fluxes~$q$ and their derivatives~$dq/d\tau$~(for details of the calculation method the reader is referred to \ref{sect:appendix_b}). The current section includes a comparison of various finite-difference scheme for solving the heat problem.

\subsection{Explicit scheme}

The problem can be solved using an explicit finite difference scheme with an embedded fixed-point iteration algorithm. The finite differences are written on a rectangular grid~$\xi_j = j/(N - 1)$, $j = 0,...,N - 1$ with a time step~$\Delta t = t_F h^2$, where~$0 < t_F \leq 1$. The discretised heat equation serves to calculate the reduced temperature~$\theta_j$ at~$j = N-2,N-3,...,0$. Let~$\widehat \theta_j$ be the temperature value at the previous timestep. The explicit scheme for the heat equation is then:

\begin{equation}
\theta_j = \widehat \theta_j + \Delta t \Lambda \widehat \theta_{j} + \frac{\Delta t \tau_0}{N_{\mathrm{P}}} \left (-\frac{d \widehat q_j}{d \tau} \right ),
\end{equation}
where the second-order differential operator is defined as~$\Lambda {\theta_j} = \left ( \theta_{j+1} - 2 \theta_{j}  + \theta_{j-1} \right )/h^2$.

The equations arising from the boundary conditions are solved iteratively:

\begin{subequations}
\begin{align}
&\stackrel{k+1}{\theta_0} = \frac{ \widehat{\theta}_1 + h \Xi - h \stackrel{k}{q_0}/N_{\mathrm{P}}}{1 + \mathrm{Bi} \cdot h}, \\
&\stackrel{k+1}{\theta_{N-1}} = \frac{ \widehat{\theta}_{N-2} + h \stackrel{k}{q_{N-1}}/N_{\mathrm{P}}}{1 + \mathrm{Bi} \cdot h},
\end{align}
\end{subequations}
where~$\Xi$ is the pulse function, $k$ is the iteration number. 

\subsection{Implicit schemes}
\label{sect:implicit_fdm}

Both the fully-implicit and semi-implicit scheme follow the same solution logic~\cite{Lunev2020}. The discrete heat equation is written as follows~\cite{samarskii2001theory}:
\begin{align}
&\frac{\stackrel{k+1}{\theta_j} - \widehat \theta_j}{\Delta t} = \left ( \sigma \Lambda \stackrel{k+1}{\theta_j} + (1 - \sigma) \Lambda \widehat \theta_j \right ) + \frac{\tau_0}{N_{\mathrm{P}}}  \stackrel{k} \phi_i, 
\end{align}
where~$0 < \sigma \leq 1$ is the weight of the scheme~($\sigma = 1.0$ for the fully-implicit scheme); $\phi_j$ is some finite-difference representation of the term~$-{d q}/{d t}$. 

After some elementary algebra, the following expressions are derived, completing the difference scheme:
\begin{subequations}
	\begin{align}
	&\overline \alpha_1 = \frac{2 \Delta t \sigma}{h^2 + 2\Delta t \sigma(1 + h \mathrm{Bi})}, \\
	&\overline \beta_1 = \frac{2 \Delta t h (\sigma \Xi + (1 - \sigma) \widehat \Xi - \sigma q_0/N_{\mathrm{P}} - (1 - \sigma) \widehat q_0/N_{\mathrm{P}} ) }  {h^2 + 2\Delta t \sigma(1 + h \mathrm{Bi})}   + \nonumber \\
	&\frac{h^2(\widehat \theta_0 + \phi_0 \Delta t) + 2\Delta t(1 - \sigma) \left [ \widehat \theta_1 - \widehat \theta_0 (1 + \mathrm{Bi} \cdot h) \right ]}{h^2 + 2\Delta t \sigma(1 + h \mathrm{Bi})}, \\
	&\theta_{N-1} = \left \{ \sigma \overline \beta_{N-1} + \frac{h^2}{2 \Delta t} \widehat \theta_{N-1} + 0.5 h^2 \phi_{N-1} + \right . \nonumber \\
	&\left . (1 - \sigma) \left [ \widehat \theta_{N-2} - \widehat \theta_{N-1} (1 + h \mathrm{Bi}) + \frac{h}{N_{\mathrm{P}}} ( \sigma q_{N-1} + (1 - \sigma) \widehat q_{N-1} ) \right ] \right \} \nonumber \\
	&/ \left \{ \frac{h^2}{2 \Delta t} + \sigma (1 + h \mathrm{Bi} - \alpha_{N-1} ) \right \}.
	\end{align}
\end{subequations}

The scheme is solved iteratively until converged values of~$\stackrel{k+1}{\theta_0}$ and~$\stackrel{k+1}{\theta_{N-1}}$ are obtained~(usually a few iterations are required). It is at least~$O(h^4 + \Delta t^2)$ accurate if~\cite{samarskii2001theory}:

\begin{subequations}
\begin{align}
&\phi_j^i = \frac{5}{6} \dot{q}_j^{i+1/2} + \frac{1}{12} \left ( \dot{q}_{j-1}^{i+1/2} + \dot{q}_{j+1}^{i+1/2} \right ), \\
&\sigma = \frac{1}{2} - \frac{h^2}{12 \Delta t},
\end{align}
\end{subequations}
where the superscript indicates averaging over two consequent time steps and $\dot{q} = dq/d\tau$.

\subsection{Verification and benchmarking}

\begin{figure}[!ht]
	\centering
	\includegraphics{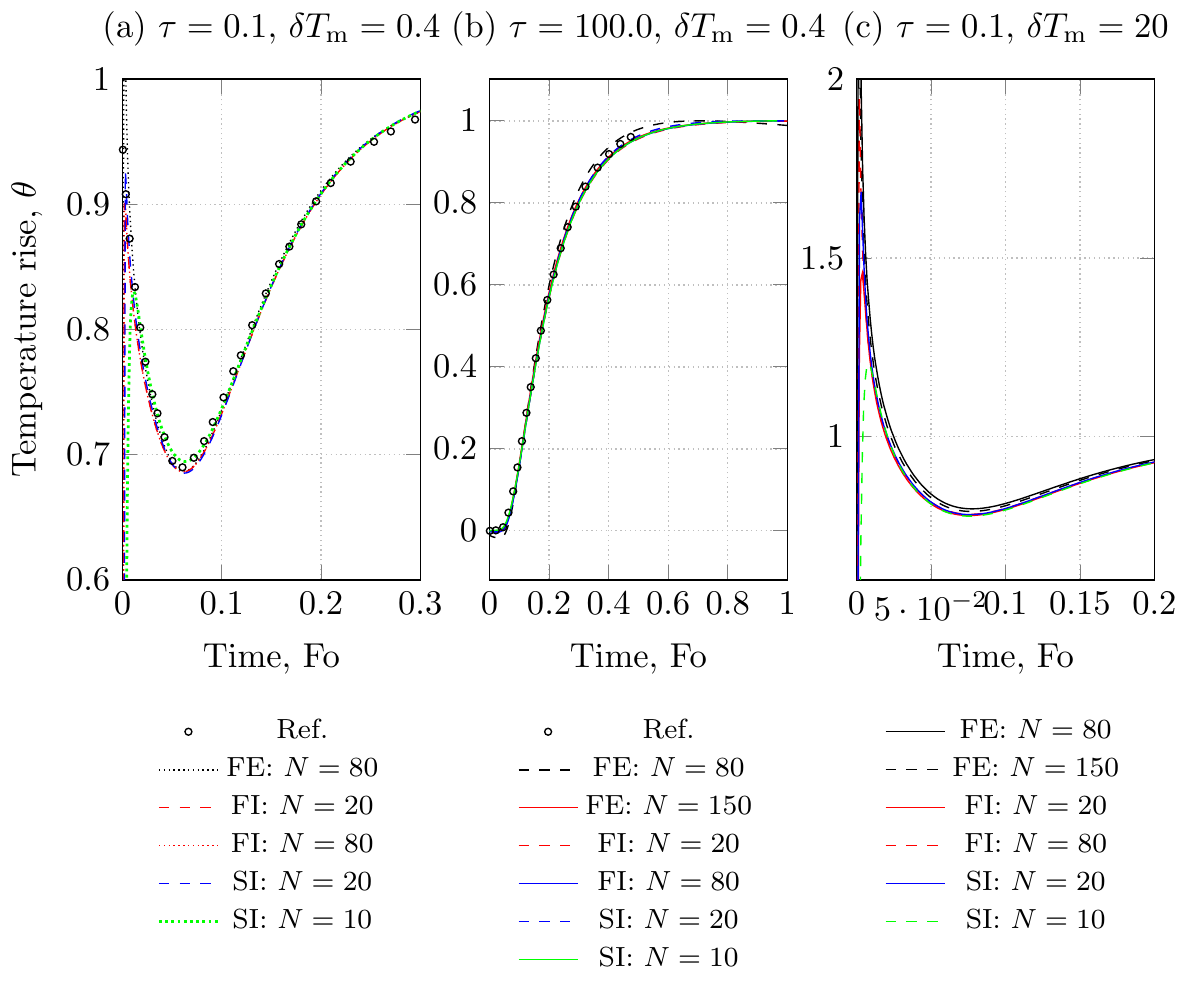}
	\caption{\label{fig:analytical_comparison} Comparison of calculation results using a fully-explicit~(FE), fully-implicit~(FI) and semi-implicit fourth-order~(SI) difference schemes with the reference analytical solutions obtained for a linearised heat equation~(digitised graphs from~\cite{Andre1998}). The analytical solution is exact at~$\delta T_{\mathrm{m}}/T_0 = 0.4/800$ but becomes invalid at higher~$\delta T_{\mathrm{m}}$ typical under experimental conditions.} 
\end{figure}

The general method~(\cref{sect:heat_problem}) and the finite-difference schemes are verified against the reference solutions reported in~\cite{Andre1998} for~$\tau_0 = 0.1$ and~$\tau_0 = 100$ at~$N_{\mathrm{P}} = 0.8612$. The calculated time-temperature profiles are shown in~\Cref{fig:analytical_comparison} where a good agreement between the linearised analytical case and the exact numerical solution is observed at~$\delta T_{\mathrm{m}} = 0.4$. Implicit schemes produce more accurate results compared to the explicit scheme; particularly, the fourth-order accurate semi-implicit scheme described in~\cref{sect:implicit_fdm} performs well even for coarse grids with~$N = 10$. This is especially important in the light of a high demand on computational resources expected when solving the inverse coupled radiative-conductive problem. It is also evident that only a numerical scheme is applicable to solving the heat problem~\cref{eq:coupled_problem} at anywhere near realistic~$\delta T_{\mathrm{m}}/T_0$ values~[note the difference between~\cref{fig:analytical_comparison} (a) and (c)].   

\section{Spatial discretisation and integration}
\label{sect:dom_method}

Commonly, the RTE~[\cref{eq:rte_dom}] is integrated using a diamond-differencing scheme~(see e.g. \cite{Jessee1997}), also known as the central-difference scheme, which for a one-dimensional problem is exactly the same as the implicit trapezoidal rule -- a second-order accurate and A-stable method. All alternative conventional methods are based on the finite-volume methodology~\cite{modest2013radiative,Liu1996} and include: the first-order step scheme, the second-order exponential, hybrid and CLAM schemes. Advances in spatial discretisation schemes for RTE, mainly based on NVD and TVD for multi-dimensional radiative transfer, have been reviewed in~\cite{Coelho2014} -- however, with no significant progress reported for high-order spatial differencing schemes. More recently, \citet{Maginot2016} have used a stiffly-accurate single diagonally implicit Runge-Kutta~(SDIRK) method reported originally by~\citet{Alexander1977}. Notable implementations of SDIRK are included in~\cite{Alessandro2018,Boom2018}. Despite their advantages, SDIRK methods only allow a stage-order of one~\cite{hairer1996stiff}. Higher stage-order is useful since this strongly improves accuracy when applied to stiff problems and increases the error-estimate quality~\cite{kennedy2016diagonally}. Stage-order two may be achieved with the first-stage explicit SDIRK~(ESDIRK). Alternative to ESDIRK is the Rosenbrock method~\cite{hairer1996stiff}, which might be more efficient for some problems~\cite{blom2016comparison}.       

\subsection{Explicit Runge-Kutta with an adaptive uniform grid}
\label{sect:rk_explicit}

A given ODE can have varying stiffness depending on the parameter values. For the RTE, stiffness is mainly determined by~$\tau_0$. At~$\tau_0 < 1$ the problem can be effectively treated as non-stiff. When stiffness is not an issue, explicit embedded Runge-Kutta schemes can be used. If high accuracy is desired, a good fourth-order scheme such as the Dormand-Prince~(DP54)~\cite{DormanPrince54} scheme with a fifth-order error control and an extended region of absolute stability can be used. Practice shows that for the current use of the RTE, error tolerance can be high, thus a lower order embedded method might be sufficient. In this case, a third-order Bogacki-Shampine~(BS32)~\cite{BS32} scheme with second-order error control and good stability can be used. Both schemes are FSAL~(first same as last), which saves computational time, and their implementation follows the same pattern described below.

Firstly, let $h_l$ denote the signed grid step, which is positive when approaching the right boundary and negative otherwise. The following notations are used: the intensities at each stage~$n = 1,...,s$ are denoted as~$i^{(n)}$ with the corresponding coordinate~$t^{(n)}_{l} = t_{l} + h_l c_{n}$, where~$m$ and~$l$ stand for the angular and spatial indices respectively. Matrix elements of the Butcher tableau are denoted as~$a_{nn'}$, and $b_n$ are the coefficients at the final stage~$s$ corresponding to~$t_l^{(s)} = t_{l} + h_l$, such that $i_{ml+1} := i_m^{(s)} = i_{ml} + h_l \sum_{n=1}^s{f^{(n)}_m b_n}$. Additionally, $\widehat{b}_n$ are the components of the error estimator. First stage is either copied from the last stage of the previous step~(if available) or calculated using the derivative~$f_{ml} := f_m^{(0)}$ at~$\tau = t_{l}$. 

The derivative at any stage~$n = 1,...,s$ is expressed e.g. for the left-to-right sweep:

\begin{align}
\label{eq:explicit_derivatives}
&f_{m}^{(n)}  = \frac{1}{\mu_m} \left (-i^{(n-1)}_{m} \left [1 - \frac{\omega_0}{2} w_m \Phi_{mm} \right ] + (1 - \omega_0) j \left ( t^{(n)}_{l} \right )   \right . \nonumber \\
&\left . + \underbrace{\frac{\omega_0}{2} \sum_{m' \neq m}^{\mu_{m'} > 0}{i_{m'}^{(n-1)} \Phi_{mm'} w_{m'}}}_{\mathrm{outward}} + \underbrace{\frac{\omega_0}{2} \sum_{m' \neq m}^{\mu_{m'} < 0}{i_{m'l+c_n}} \Phi_{mm'} w_{m'}}_{\mathrm{inward}} \right ).
\end{align}      
where $i^{(n)}_m =  i_{ml} + h_l \sum_{n'=1}^{n - 1}{a_{nn'} f_{m}^{(n')}}$ are the outward intensities at the node~$m$ and stage~$n$. Depending on whether the RTE is solved left-to-right or right-to-left, the angular index~$m$ for the outward intensities will run through the indices of either positive or negative nodes (cosines). For the sum over outward intensities, the latter are expressed in the same way using the solution~$i_{m'}^{(n-1)}$ at the stage~$n-1$. Inward intensities~$i_{m'l+c_n}$ are not known \textit{a priori}, which is why the RTE is solved iteratively; this will be described in more detail later in the text. For now these intensities are assumed to be known.

Once the derivative~$f_m^{(n)}$ becomes known, it is then used to calculate the next stage approximation~$i^{(n)}_m$, and so on. This process repeats for all~$n = 1,...,s$. As soon as all derivatives have been calculated, the intensities~$i_{ml+1}$ may be evaluated using the respective expression. Error control is achieved by evaluating the vector~$\mathbf{est}$. the components of which are given by:

\begin{align}
\label{eq:est}
\mathrm{est}_m = h \sum_{n = 1}^{s} { (b_n - \widehat{b}_n) f_{m}^{(n)} },
\end{align} 
where~$m$ runs through the indices of outward intensities. 

Absolute and relative tolerances are introduced according to~\citet{hairer1993solving} so that the error threshold is defined via:

\begin{align}
\label{eq:error_control}
e_{l\pm1} = atol + \max_m {\left (|i_{ml}|, |i_{ml\pm1}| \right )} \times rtol.
\end{align}

Thus, $\max_m {(\mathrm{est}_m)}$ is compared at each subsequent integration step $l\pm1$ against~$e_{l\pm1}$ -- if the former is greater than the latter, integration stops immediately, triggering a grid re-construction with a different segmentation:~$N^{[u+1]} = s_G N^{[u]}$~(typically~$s_G = 1.5$), where~$[u]$ indicates the value at current iteration. 

As mentioned above, to solve the RTE, one must calculate the intensities corresponding to both the negative and positive~$\mu_m$. However, when using the method above to solve either of the Cauchy problems, only half of the intensities is readily calculated while the other half is assumed to be known. To solve the RTE for all~$\mu_m$, an iterative solution is required. Here two techniques are considered~\cite{demmel1997applied}: the fixed-point iterations and the successive over-relaxation. In both cases, the intensities at iteration~$[u + 1]$ are expressed as:

\begin{align}
i^{[u+1]}_{ml} = (1 - \omega_R) i^{[u]}_{ml} + \omega_R i_{ml},
\end{align}   
where the relaxation parameter~$\omega_R = 1$ for fixed-point iterations and~$1 < \omega_R < 2$ in the successive over-relaxation technique. The second term on the right-hand side is the solution of the ODEs times the relaxation parameter. For instance, at~$\omega_R = 1.7$ and for pure isotropic scattering at~$\omega_0 = 1$, convergence is reached two times faster than for fixed-point iterations.

The stopping criterion for the iterative procedure regards the relative change to the boundary fluxes~$q_0$ and~$q_N$ at the left and right boundaries correspondingly:

\begin{align}
\frac{\left | q_0^{[u + 1]} - q_0^{[u]} \right | + \left | q_{N}^{[u + 1]} - q_{N}^{[u]} \right |}{ \left | q_0^{[u]} + q_{N}^{[u]} \right | } < e_{\mathrm{it}},
\end{align}  
where~$e_{\mathrm{it}}$ is a relative error tolerance~(typically, $e_{\mathrm{it}} \simeq 10^{-4}$). 

\subsection{TR-BDF2 with an adaptive stretching grid}
\label{sect:esdirk}

For moderately- and highly-stiff problems, e.g. at~$\tau_0 > 10$, the use of a uniform grid requires a very small step size~$h_l$ to make the scheme stable, thus greatly increasing the computational cost of an explicit method. Hence, an adaptive step-size control should be used instead, which achieves true flexibility in the~$A/L$-stable, stiffly-accurate methods, for instance, the TR-BDF2 scheme~\cite{TRBDF2}. The latter can be regarded as a major improvement over the original diamond-differencing scheme for plane-parallel radiative transfer problems, since it includes the same trapezoidal rule~(diamond-differencing) at the second stage and uses second-order backward-differencing at the third stage, resulting in stiff accuracy. Furthermore, it provides an asymptotically correct error estimate and allows dense output. TR-BDF2 can be regarded as an ESDIRK scheme~\cite{kennedy2016diagonally}.

The explicit first stage is calculated in the same way as in~\cref{sect:rk_explicit}, noting that TR-BDF2 is also FSAL. The second and third stages are implicit by definition. However, because the ODEs in the DOM are linear, the corresponding intensities can easily be found explicitly from the solution of the following linear set. For instance, the left-to-right sweep at the second stage:

\begin{align}
\label{eq:implicit_derivative}
&i^{(2)}_{m} \left [ 1 + \frac{h_l d}{\mu_m} \left ( 1 - \frac{\omega_0}{2} w_m \Phi_{mm} \right ) \right ] - \frac{h_l d}{\mu_m} \underbrace{\frac{\omega_0}{2} \sum_{m' \neq m}^{\mu_{m'} > 0}{i^{(2)}_{m'}} \Phi_{m'm} w_{m'}}_{\mathrm{outward}} = \nonumber \\
&i_{ml} + h_l d f_m^{(1)} + \frac{h_l d}{\mu_m} \left [ (1 - \omega_0) j(t_l + \gamma h_l) + \underbrace{\frac{\omega_0}{2} \sum_{m' \neq m}^{\mu_{m'} < 0}{i_{m'l+\gamma}} \Phi_{m'm} w_{m'}}_{\mathrm{inward}}   \right ], 
\end{align}
where~$\gamma = 2 - \sqrt{2}$ and~$d = \gamma/2$~\cite{TRBDF2}.

Clearly, this reduces to a linear matrix equation~$\mathbf{A} \mathbf{i}^{(2)}_{ml+\gamma} = \mathbf{B}^{(2)}_{ml+\gamma}$, which is solved by matrix inversion. Due to the~$\mathbf{A}$ matrix usually being low-dimensional~(the dimension is equal to a half of the total number of quadrature points), a fast matrix inversion routine has been implemented for the typical quadrature sets. For higher-order quadratures, a matrix inversion tool based on either QR, LU or Cholesky decomposition of the Apache Commons Mathematics Library is used. Since the method is E\textbf{S}DIRK, the final third stage uses the same matrix inverse~$\mathbf{A}^{-1}$. The linear set for the third~(and final) stage is~(left-to-right sweep):

\begin{align}
&\mathbf{A} \mathbf{i}^{(3)}_{ml+1} = i_{ml} \left ( 1 - \frac{w}{d} \right ) + \frac{w}{d}i_{m}^{(2)} + \nonumber \\ 
&\frac{h_l d}{\mu_m} \left [ (1 - \omega_0) j(t_l + h_l) + \underbrace{\frac{\omega_0}{2} \sum_{m' \neq m}^{\mu_{m'} < 0}{i_{m'l+1}} \Phi_{mm'} w_{m'}}_{\mathrm{inward}}   \right ], 
\end{align}    
where~$w = \sqrt{2}/4$~\cite{TRBDF2} (this should not be confused with the quadrature weights~$w_m$).

The correct error estimate~\cite{TRBDF2} valid for both stiff and non-stiff problems is then simply: $\mathbf{Est} = \mathbf{A}^{-1} \mathbf{est}$, where $\mathbf{est}$ is given by~\cref{eq:est} and~\cite{TRBDF2}:

\begin{equation}
\widehat{\mathbf{b}}^{\mathbf{T}} = \left( \ (1 - w)/3, \ (3w + 1), \ d/3 \ \right).
\end{equation}

The same general scheme for error control~[\cref{eq:error_control}] is used. 

To take advantage of the stability properties of TR-BDF2, an adaptive grid is constructed using stretching functions~\cite{Stretching1,Stretching2}. Since rapid variation of intensities is mainly expected when approaching the boundaries, it is sufficient to maintain a small step in their vicinity. The stiff solver can then use an arbitrary large step in the remainder domain. For this purpose, the grid step~$h_l$ is defined via a hyperbolic tangent function:

\begin{align}
\label{eq:stretched_grid}
h_l = \frac{\tau_0}{2} \left [ 1.0 - \frac{\tanh \left \{ S_g (1 - 2 \xi_l) \right \} }{\tanh(S_g)} \right ],	\quad \xi_l = 1.0/N^{[u]}_g,
\end{align}
where~$N^{[u]}_g$ is the number of segments in a uniform grid and $S_g$ is the stretching factor. 

\begin{figure}[h]
	\centering
	\includegraphics{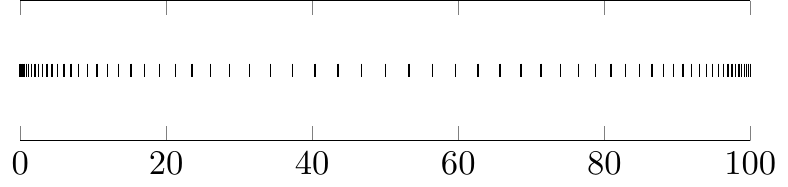}
	\caption{A symmetric stretched grid generated using \cref{eq:stretched_grid} at~$N^{[u]}_g = 64$, $S_g = 3.0$ and~$\tau_0 = 100.0$}
	\label{fig:stretched_grid}
\end{figure}

\Cref{fig:stretched_grid} shows an example grid generated using the above algorithm. 

When the error becomes higher than the threshold given by~\cref{eq:error_control}, the grid is re-constructed by increasing the number of grid points in the same manner as described in \cref{sect:rk_explicit}. The first iteration always starts from a uniform grid with a default of~$N^{[u]}_g = 8$ segments. The parameter~$S_g$ normally does not change during the re-construction. Finally, the same iterative procedure described in~\cref{sect:rk_explicit} is adopted to obtain convergence.   

\subsection{Interpolation}

In each case, knowledge of the dimensionless temperature~$\theta$ is required at intermediate integration steps~$t^{(n)}$ used then to calculate the reduced radiance~$j(t^{(n)})$. Since the temperature is defined discretely on a different external grid of the heat equation, an interpolation procedure is required to calculate the temperature~$\theta_l$ at the integrator nodes. In this case, the dimensionless temperature is interpolated using natural cubic splines implemented in the Apache Commons Mathematics Library. 

Both the explicit~[\cref{eq:explicit_derivatives}] and implicit~[\cref{eq:implicit_derivative}] methods contain summation over the unknown inward intensities~$i_{ml+c_n}$. Since all intensities are calculated at the internal grid points~$l$ and because~$l +c_n$ is not a grid point, an interpolation procedure is required here as well to calculate~$i^{[u]}_{ml+c_n}$ using the~$i^{[u-1]}_{ml}$ values obtained at the previous iteration. Additionally for the implicit method, the outward intensities at the intermediate points~$t_l + \gamma h_l$ are not known either, and hence the same procedure needs to be used for their calculation. Because in Runge-Kutta methods both the intensities and their derivatives are calculated, a cheap and convenient method for this interpolation is the globally~$C^1$ Hermite interpolation described in detail in~\cite{rogers1989mathematical}. The Hermite interpolant satisfying the function and derivative values at end points of the segment $t \in [a,b]$ is:

\begin{align}
h(t) = T^2 (3 - 2T) y_1 + (T-1)^2(1+2T)y_0 + \{ T^2(1-T)d_1 + (T-1)^2Td_0  \} h,
\end{align}
where $T = (t - a)/h_l$, $a = t_l$, $b = t_{l \pm 1}$, $y_0 = i_{ml}$, $y_1 = i_{ml\pm 1}$, $d_0 = f_{ml}$, $d_1 = f_{ml\pm 1}$.

This allows effective interpolation of both inward and outward intensities at any intermediate point~$0 < t < \tau_0$.

\subsection{Angular discretisation}

The quadrature choice is central to the DOM as it defines both the overall accuracy of the method and the stability requirements for the spatial integration technique. \citet{Chandrasekhar1960} originally considered the Gauss-Legendre and Lobatto~(Radau) quadratures for angular discretisation. In modern calculations, the level-symmetric quadratures by~\citet{lathrop1964discrete} are often used~\cite{modest2013radiative}. These and other similar quadratures have been reviewed in~\cite{Truelove1987,Kumar1990,Li1998,Liu2002}. More recently, an extensive review~\cite{Koch2004} of different quadratures has shown that for problems generating a continuous intensity field, the Gauss-Chebyshev quadrature LC11 derived by~\citet{Lebedev1976} offers the highest precision. Since in many cases, particularly for the one-dimensional radiative transfer with diffuse emission and reflection conditions, the intensities are discontinuous at~$\mu = 0$~(see e.g. \cite{Thynell1998}), standard quadratures which do not specifically treat the discontinuity would give inaccurate results. The level-symmetric quadratures were designed to cover both the non-continuous and discontinuous case and are applicable to a wide range of problems. However, high-order quadratures~(such as $S^{10}$, $S^{12}$ etc.) yield negative weights. Although quadratures such as~$S^8$ give sufficiently accurate results in many cases, an alternative should be considered for higher-order calculations. A composite Gaussian quadrature has been considered for Fresnel boundary conditions in~\cite{Fresnel2011} where the angular interval was divided in three segments. A similar procedure can be performed for the diffuse emission and reflection boundaries. 

Consider the~$M$ cosine nodes and weights of a Gauss-Legendre quadrature on~$[0,1]$: $\widetilde{\mu}_m$ and~$\widetilde{w}_m$. The goal is to construct a composite quadrature that will work despite the intensities being discontinuous at~$\mu = 0$. The $2M$ cosine nodes of this composite quadrature are then:

\begin{align}
\label{eq:composite_gaussian}
\mu_m = \frac{\widetilde{\mu}_m + 1}{2}, \quad \mu_{m + n/2} = -\frac{\widetilde{\mu}_m + 1}{2}. 
\end{align}
with the same weights~$w_m = \widetilde{w}_m$. 

By construction, the composite Gaussian quadrature given by~\cref{eq:composite_gaussian} is applicable to discontinuous functions at~$\mu = 0$. An example~$G^{16}_M$ ordinate set proposed in this work is given in \cref{tab:G16M}~(note this quadrature is symmetric). 

\begin{table}[]
	\centering
	\caption{Nodes~(positive half) and weights of a $G^{16}_M$ composite Gauss-Legendre quadrature for a function discontinuous at~$\mu = 0$}
	\label{tab:G16M}
	\begin{tabular}{rr}
		\hline
		\multicolumn{1}{c}{\begin{tabular}[c]{@{}c@{}}Cosine nodes, \\ $\mu_m$\end{tabular}} & \multicolumn{1}{c}{\begin{tabular}[c]{@{}c@{}}Quadrature weights, \\ $w_m$\end{tabular}} \\ \hline
		0.980144928248767                                                                       & 0.050614268145189                                                                           \\
		0.898333238706814                                                                       & 0.111190517226691                                                                           \\
		0.762766204958165                                                                       & 0.156853322938942                                                                           \\
		0.591717321247824                                                                       & 0.181341891689181                                                                           \\
		0.408282678752176                                                                       & 0.181341891689181                                                                           \\
		0.237233795041834                                                                       & 0.156853322938941                                                                           \\
		0.101666761293186                                                                       & 0.111190517226693                                                                           \\
		0.019855071751233                                                                       & 0.050614268145190                                                                           \\ \hline
	\end{tabular}
\end{table}

\subsection{Verification and benchmarking}

\begin{figure}[!ht]
	\centering
	\includegraphics{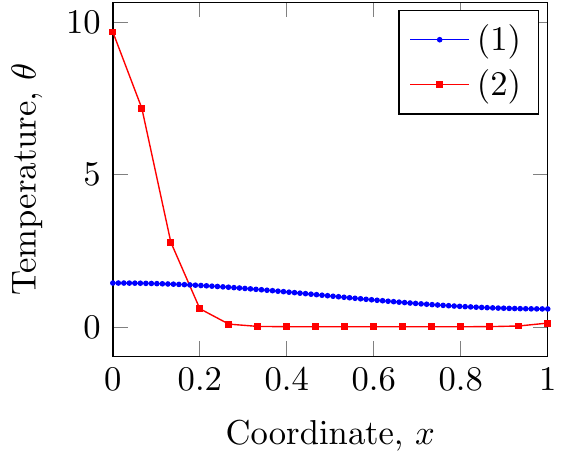}
	\caption{\label{fig:test_profile} Sample discrete dimensionless temperature profiles for verification and benchmarking purposes. The profiles are discretised differently to test the interpolation capability.} 
\end{figure}

\begin{figure}[!ht]
	\centering
	\includegraphics{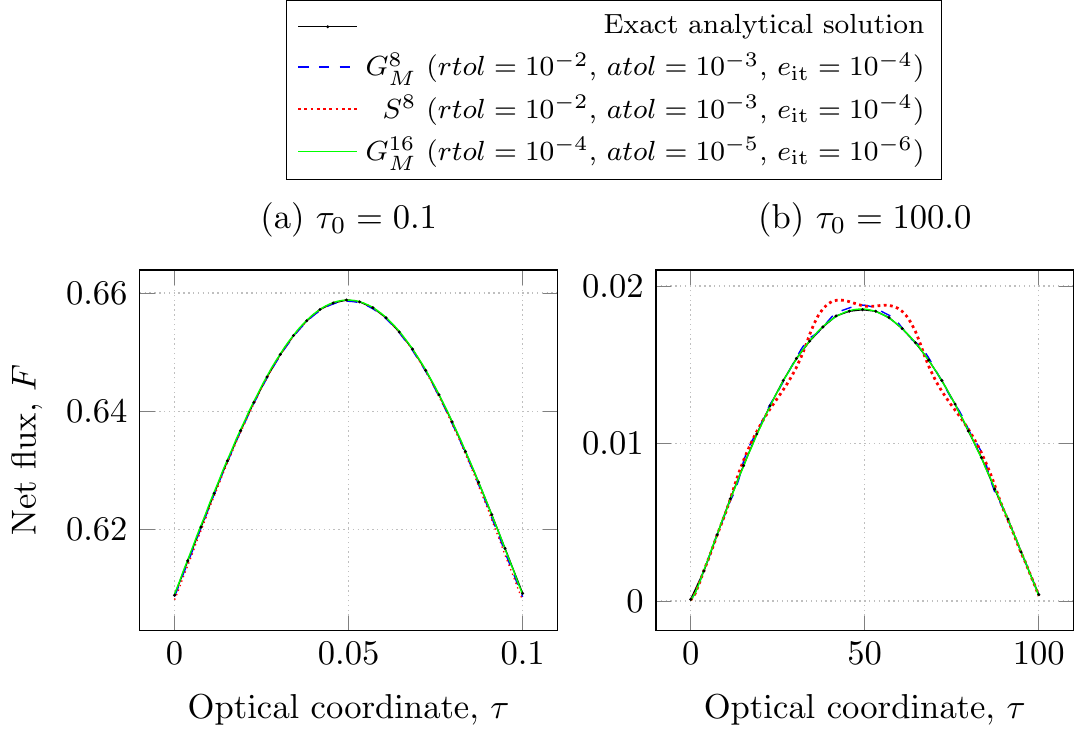}
	\caption{\label{fig:exact_comparison} Comparison of the discrete ordinates solution using a TR-BDF2 adaptive solver and different ordinate sets~($G^8_M$, $G^{16}_M$, $S^8$) and error tolerance levels with the exact analytical solution for a grey non-scattering medium~($\omega_0 = 0.0$, $\varepsilon = 0.85$).  Parameters:~$\delta T_{\mathrm{m}} = 10.0$, $T_0 = 800$~K.} 
\end{figure}

\begin{figure}[!ht]
	\centering
	\includegraphics{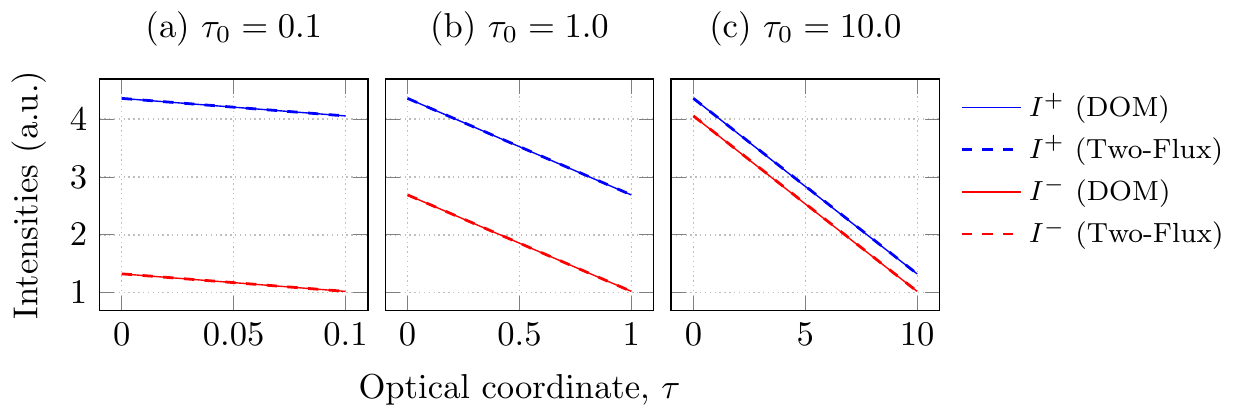}
	\caption{\label{fig:twoflux_comparison} Comparison of the discrete ordinates solution at low error tolerance using the TR-BDF2 solver and the~$G^8_M$ ordinate set with the two-flux method for pure isotropic scattering in case of black walls~($\omega_0 = 1.0$, $\varepsilon = 1.0$) for a test temperature profile (2). Parameters:~$\delta T_{\mathrm{m}} = 36.7$, $T_0 = 800$~K. The intensities have been calculated with a re-normalised radiance~$j_r = (1 + \theta {\delta T_{\mathrm{m}} }/{T_0})^4$.} 
\end{figure}

To verify the solvers and the discrete ordinate sets, two model cases were considered: \begin{enumerate*}[label=(\textit{\alph*})]
	\item a non-scattering grey medium with diffusely emitting and reflecting walls~($\varepsilon = 0.85$, $\omega_0 = 0.0$);
	\item an isotropic perfectly scattering medium with black walls~($\varepsilon = 1.0$, $\omega_0 = 1.0$) 
\end{enumerate*}. In the first case~(\cref{fig:exact_comparison}), the DOM solution was compared against an exact analytical solution, whereas the second comparison~(\cref{fig:twoflux_comparison}) was made in reference to the two-flux model~\cref{sect:two-flux}. The equations were solved using the GNU Octave/Matlab bvp5c solver. Two temperature profiles were used -- both are shown in~\cref{fig:test_profile}. The parameter~$\tau_0$ was allowed to vary from~$\tau_0 = 0.1$~(non-stiff) to~$\tau_0 = 100.0$~(very stiff).

Results for the three quadratures considered~($G^8_M$, $S^8$, $G^{16}_M$) show good overall agreement, with the~$G^8_M$ and $G^{16}_M$ quadrature producing significantly less deviation from the reference analytic solution~(\cref{fig:exact_comparison}) at the boundaries~(non-stiff case) and at intermediate points~(stiff case). The deviation is decreased even more when a low error tolerance is selected~($rtol = 10^{-4}$, $atol = 10^{-5}$, $e_{\mathrm{it}} = 10^{-6}$). For comparison with the two-flux model, an artificial quadrature containing two equal-weight symmetric points is examined. An exact match between the approximate analytical model and the discrete ordinates method is shown in~\cref{fig:twoflux_comparison}, thus confirming the reliability of the numeric procedure.


\begin{figure}[!ht]
	\centering
	\includegraphics{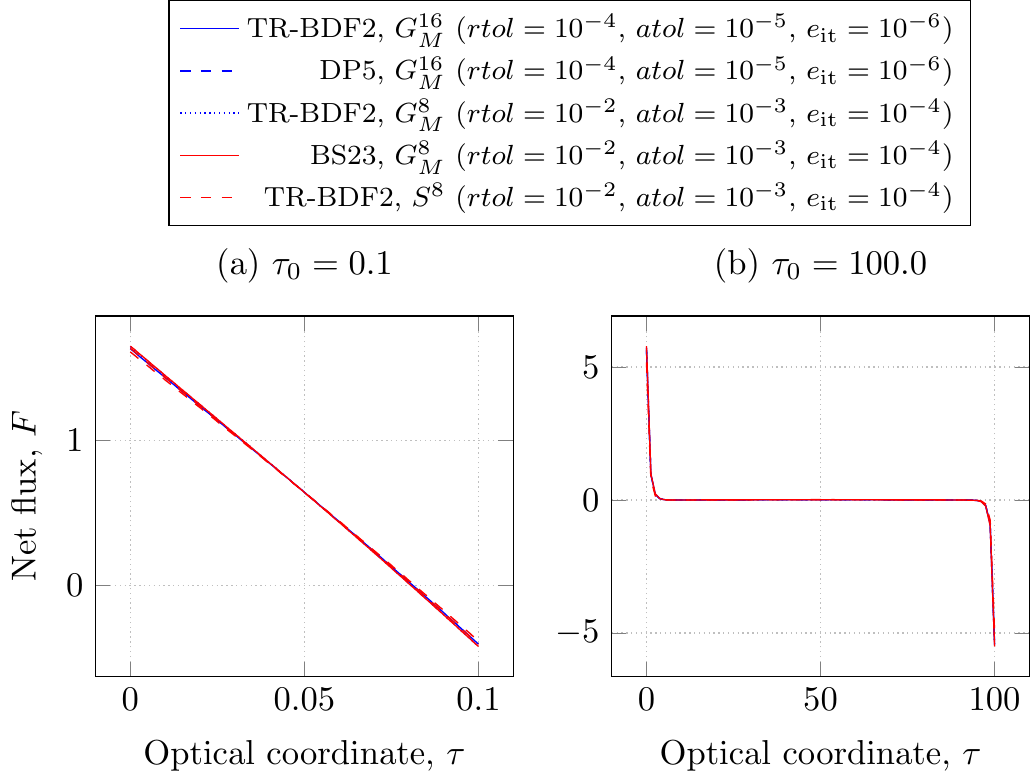}
	\caption{\label{fig:anisotropic_comparison} Comparison between the discrete ordinates solutions using different ordinate sets~($G^8_M$, $G^{16}_M$, $S^8$) and solvers~(BS32, DP5, TR-BDF2) at different error tolerance levels for a grey medium with a strong anisotropic scattering~($\varepsilon = 0.85$, $\omega_0 = 0.4$, $g = 0.8$). Parameters:~$\delta T_{\mathrm{m}} = 10.0$, $T_0 = 800$~K.} 
\end{figure}

Additionally, the performance of different schemes and quadratures was tested for a grey medium with a strong anisotropic scattering~($\varepsilon = 0.85$, $\omega_0 = 0.4$, $g = 0.8$). The results of different computational methods for the net fluxes shown in~\cref{fig:anisotropic_comparison} show good mutual agreement both in the stiff and non-stiff cases.

Finally, the relative performance of different schemes was assessed in~\cref{tab:solvers_benchmark}. Here the TR-BDF2 scheme in the high-tolerance mode using the~$G^8_M$ quadrature was used as reference, corresponding to the respective~$1.00$ table entry. Increasing problem stiffness in the high-tolerance mode only marginally increases the computational cost for TR-BDF2. Other schemes do not perform so well in terms of performance, particularly the DP5 at~$\tau_0 = 100$ is 50 times slower than the reference. BS23 performs better but still fails to deliver a reasonable computation time for stiff problems. For the~$G^{16}_M$ quadrature there was no fast matrix inversion implemented and hence the TR-BDF2 algorithm relied on a generic decomposition algorithm for the latter. This justifies the considerably more expensive calculations. Problems requiring only a small ordinates set~(e.g. $S^4$) show a~$\approx 1.5$ increase in performance compared to the reference. Same performance for~TR-BDF2 and DP5 is achieved at low error tolerance levels for a non-stiff~($\tau_0 = 0.1$) problem, whereas~BS23 requires a finer step size, which almost triples the overall cost. The numbers change dramatically even for moderately-stiff problems~($\tau_0 = 10.0$), with the DP5 outperforming the BS23 scheme -- as expected, since DP5 is a fourth-order method. On the other hand, both require more resources to achieve the same error tolerance compared to the TR-BDF2 due to the adaptive grid employed for the latter. For the~$G^{16}_M$ quadrature there is an expected drop in performance -- and vice versa for the~$S^4$ ordinate set. 

With these results in mind, the default settings for calculation are chosen as TR-BDF2 and a $G^8_M$ ordinate set in the high-tolerance mode.

\begin{table}[]
	\centering
	\caption{Benchmark results of the DOM for high~($rtol = 10^{-4}$, $atol = 10^{-5}$, $e_{\mathrm{it}} = 10^{-6}$) and low~~($rtol = 10^{-2}$, $atol = 10^{-3}$, $e_{\mathrm{it}} = 10^{-4}$) error tolerance levels using a test temperature profile at~$T_0 = 800$~K and~$\delta T_{\mathrm{m}} = 10.0$ for strong anisotropic scattering in a grey medium~($\omega_0 = 0.4$, $g = 0.8$, $\varepsilon = 0.85$)}
	\label{tab:solvers_benchmark}
	\begin{tabular}{@{}cllrrr@{}}
		\toprule
		\multicolumn{1}{l}{\multirow{2}{*}{Error tolerance}} & \multicolumn{1}{c}{\multirow{2}{*}{Quadrature}} & \multicolumn{1}{c}{\multirow{2}{*}{Solver}} & \multicolumn{3}{c}{Computational cost (rel.)}                                                             \\ \cmidrule(l){4-6} 
		\multicolumn{1}{l}{}                                 & \multicolumn{1}{c}{}                            & \multicolumn{1}{c}{}                        & \multicolumn{1}{c}{$\tau_0=0.1$} & \multicolumn{1}{c}{$\tau_0=10.0$} & \multicolumn{1}{c}{$\tau_0=100.0$} \\ \midrule
		\multirow{5}{*}{High}                                & $G^8_M$                                         & TR-BDF2                                     & 1.00                             & 1.00                              & 1.90                               \\
		&                                                 & DP5                                         & 1.55                             & 6.80                              & 50.0                               \\
		&                                                 & BS23                                        & 1.04                             & 4.90                              & 28.1                               \\
		& $G^{16}_M$                                      & TR-BDF2                                     & 3.76                             & 4.95                              & 20.0                               \\
		& $S^4$                                           &                                             & 0.67                             & 0.64                              & 0.83                               \\ \midrule
		\multirow{5}{*}{Low}                                 & $G^8_M$                                         & TR-BDF2                                     & 1.00                             & 4.1                               & 9.86                               \\
		&                                                 & DP5                                         & 1.54                             & 19.0                              & 193.4                              \\
		&                                                 & BS23                                        & 2.84                             & 110.5                             & -                                  \\
		& $G^{16}_M$                                      & TR-BDF2                                     & 13.64                            & 23.14                             & 146.62                             \\
		& $S^4$                                           &                                             & 0.85                             & 2.30                              & 2.475                              \\ \bottomrule
	\end{tabular}
\end{table}

\section{Cross-verification}

\begin{figure}
	\centering
	\includegraphics{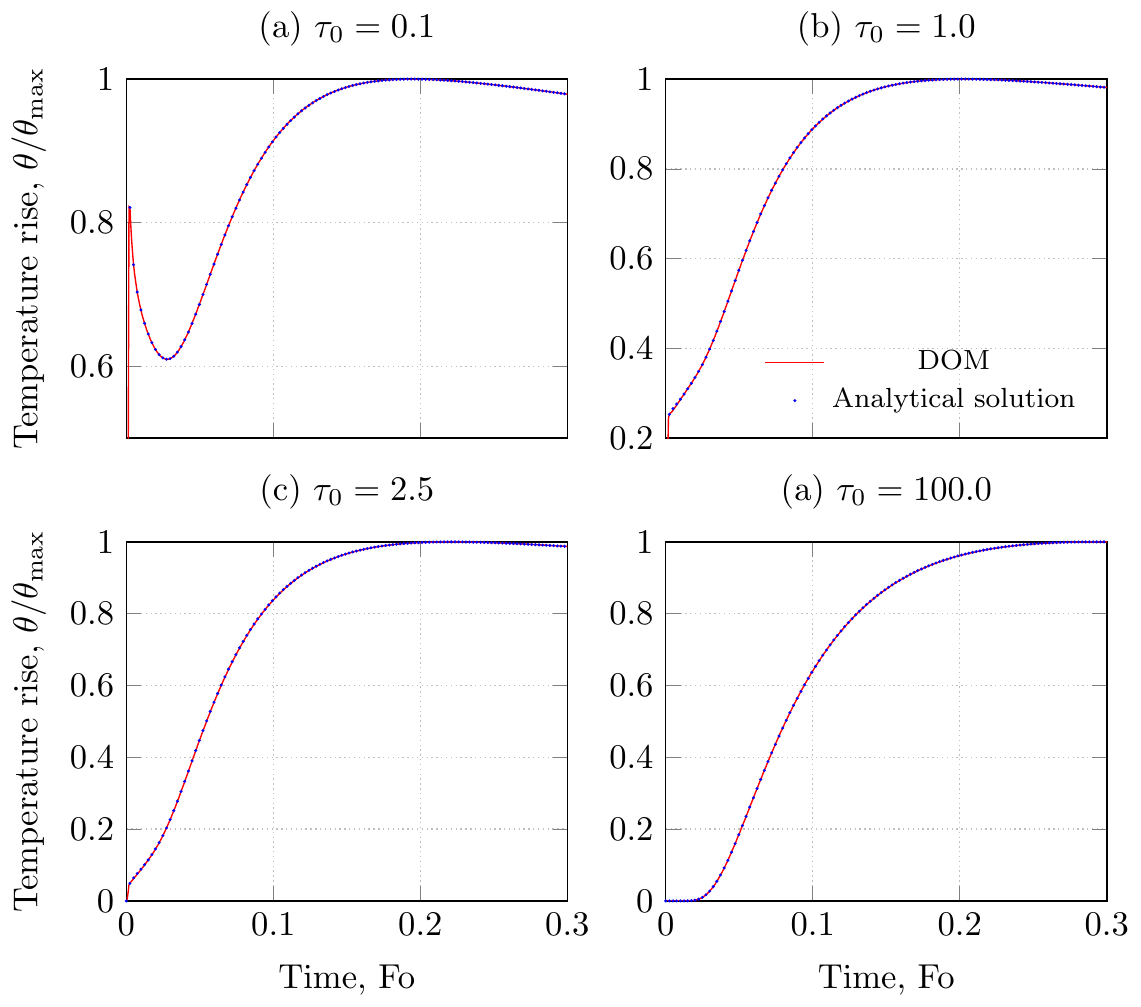}
	\caption{\label{fig:cross-verification} Finite-difference~(4th order SI scheme, $N=16$) solutions to the heat problem with a non-scattering radiative transfer calculated using either DOM~($G^8_M$, TR-BDF2) or the analytical formula where a four-point Chandrasekhar's quadrature is used.} 
\end{figure}

The goal is to verify the complete solution to the conductive-radiative problem described in \cref{sect:schemes,sect:dom_method}. Synthetic model parameters used in the tests are listed in~\cref{tab:verification_parameters}. These correspond to a case of non-scattering grey medium; the latter is especially helpful since it allows an exact solution to the RTE~(\cref{sect:rte_exact}), examples of which have previously been shown in~\cref{fig:analytical_comparison}. The resulting time-temperature profiles generated by solving the boundary problem~[\cref{eq:coupled_problem}] with the radiative fluxes calculated using the discrete ordinates method were compared to the same profiles calculated using the analytical solution to the RTE. No deviation between the two calculation methods is observed~(\Cref{fig:cross-verification}), thus indicating a correct implementation of all solvers.

\begin{table}[!h]
\centering
\caption{Test calculation parameters}
\label{tab:verification_parameters}
\begin{tabular}{@{}llrl@{}}
\toprule
\multicolumn{1}{c}{Parameter} & \multicolumn{1}{c}{Notation} & \multicolumn{1}{c}{Value} & \multicolumn{1}{c}{Units}                         \\ \midrule
Planck number                 & $N_{\mathrm{P}}$             & 0.8612                    &                                                   \\
Scattering albedo             & $\omega_0$                   & 0.0                       &                                                   \\
Biot number                   & Bi                           & 0.1                       &                                                   \\
Test temperature              & $T_0$                        & 1486                      & K                                                 \\
Laser energy                  & $Q_{\mathrm{las}}$           & 5                         & J                                                 \\
Specific heat                 & $C_{\mathrm{p}}$             & 1296                      & $\mathrm{J} \ \mathrm{kg}^{-1} \ \mathrm{K}^{-1}$ \\
Density                       & $\rho$                       & 3735                      & kg m$^{-3}$                                       \\
Thermal diffusivity           & $a$                          & 1.254                     & $\mathrm{mm}^2 \mathrm{s}^{-1}$                   \\
Pulse width                   & $t_{\mathrm{las}}$           & 1.5                       & ms                                                \\
Thickness                     & $l$                          & 1                         & mm                                                \\
Diameter                      & $d$                          & 10                        & mm                                                \\ \bottomrule
\end{tabular}
\end{table}

\section{Experimental validation}

\begin{figure}
	\centering
	\includegraphics{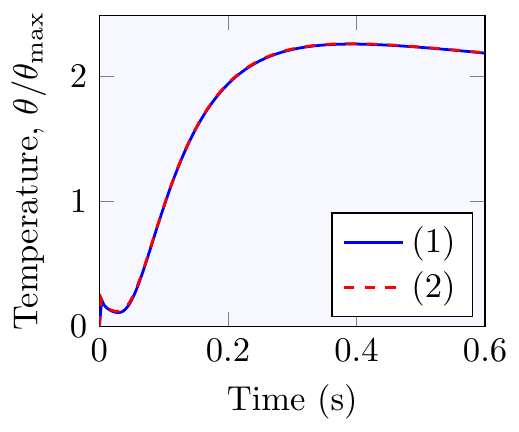}
	
	\begin{tabular}{@{}lll@{}}
		\toprule
		\multicolumn{1}{c}{\multirow{2}{*}{Parameter}} & \multicolumn{2}{c}{Value}                         \\ \cmidrule(l){2-3} 
		\multicolumn{1}{c}{}                           & \multicolumn{1}{c}{(1)} & \multicolumn{1}{c}{(2)} \\ \midrule
		Bi                                             & 0.1153                  & 0.11071                 \\
		$a$ (mm$^2$s$^{-1})$                            & 1.5314                  & 1.48182                 \\
		$\omega_0$                                     & 0.81557                 & 0.76694                 \\
		$g$                                            & 0.93998                 & 0.22385                 \\
		$N_{\mathrm{P}}$                               & 25.48625                & 6.48487                 \\
		$\tau_0$                                       & 0.29264                 & 1.55839                 \\ \bottomrule
	\end{tabular}
	\caption{\label{fig:ill_posed} Two seemingly equal solutions based on completely different parameter sets. The $N_{\mathrm{P}}$ parameter value in Set~(1) leads to a physically impossible refractive index~$n=0.75$, whereas Set (2) yields~$n=1.48$.} 
\end{figure}

Experiments conducted with the use of a laser flash analyser~(LFA) produce raw data in the form of time-temperature profiles with varying level of noise~\cite{Lunev2020}. Experimental validation requires solving the inverse problem of heat transfer, which boils down to finding a set of parameters~(e.g. \cref{tab:verification_parameters}) corresponding to an optimal solution of the heat problem. A solution is deemed optimal if the objective function~(such as the sum of squared residuals) reaches a global minimum in the parameter space. Fortunately, the corresponding optimisation procedure has already been previously implemented and extensively tested in~\cite{Lunev2020}. Nevertheless, some modifications to the procedure are required both for the diathermic model~(\cref{sect:diathermic}) and the coupled conductive-radiative problem~(\cref{sect:heat_problem}). Firstly, the original linear-interpolation procedure has been replaced by spline interpolation. Secondly, the basic procedure in~\cite{Lunev2020} involved only unconstrained optimisation. In case of an ill-posed problem or a tendency of the computational method to fail outside a certain region in the parameter space, the unconstrained optimisation procedure will not behave well. \Cref{fig:ill_posed} shows two almost identical time-temperature profiles obtained with two very different parameter sets. This is a classical example of an ill-posed problem~\cite{tikhonov1963solution}. To eliminate non-physical solutions, the parameter space should be bounded. The corresponding linear constraints are listed in~\cref{tab:parameter_bounds}. The complete solution of the optimisation problem with linear constraints based on the active-set method has been discussed in~\cite{gill2019practical} and the general method of solving ill-posed problems is known as the Tikhonov regularisation. A very simple alternative is considered in this work mainly for demonstration purposes. A one-to-one mapping~$Y_i \in \mathbb{R} \to X_i \in [a,b]$ is introduced for each parameter~$y_i$ in~\cref{tab:parameter_bounds} using hyperbolic functions. This ensures that at each time the parameter~$x_i$ only takes `reasonable' values. The optimisation procedure is then effectively the same, except that the search vector is formed of~$X_i$ rather than~$Y_i$. It should also be noted that imposing these constrains is only possible if the thermal properties of the sample~(specific heat and density) are known in each experiment -- otherwise there is no way of telling whether the parameter value is sensible or not. As a direct consequence, this means that even the diathermic model, which does not require neither the specific heat nor the density values for calculation, will not guarantee physically reasonable results if the thermal properties are unknown and an unconstrained optimisation is used instead.        

\begin{table}[]
	\centering
	\caption{Parameter bounds and their one-to-one monotonic mapping}
	\label{tab:parameter_bounds}
	\begin{tabular}{lll}
		\hline
		\multicolumn{1}{c}{Parameter, $Y_i$} & Bounds                                                                                                                                              & Mapping                                                              \\ \hline
		Bi                                   & $0 \leq \mathrm{Bi} \leq (4\sigma_0 T^3 l)/\lambda$                                        & \multirow{3}{*}{$Y_i = 0.5 Y_i^{\max} \left ( 1 + \tanh (X_i) \right )$} \\
		$N_{\mathrm{P}}$                      & $0 < N_{\mathrm{P}} \leq \lambda (4\sigma_0 T^3 l)^{-1}$ &                                                                             \\
		$\omega_0$                           & $0 \leq \omega \leq 1$                                                                                                                              &                                                                             \\
		$g$                                  & $-1 \leq g \leq 1$                                                                                                                                  & $Y_i = \mathrm{tanh}(X_i)$                                                             \\
		$\tau_0$                             & $\tau_0 > 0$                                                                                                                                        & $Y_i = e^{X_i}$                                                            \\ \hline
	\end{tabular}
\end{table}

\begin{figure}
	\centering
	\includegraphics{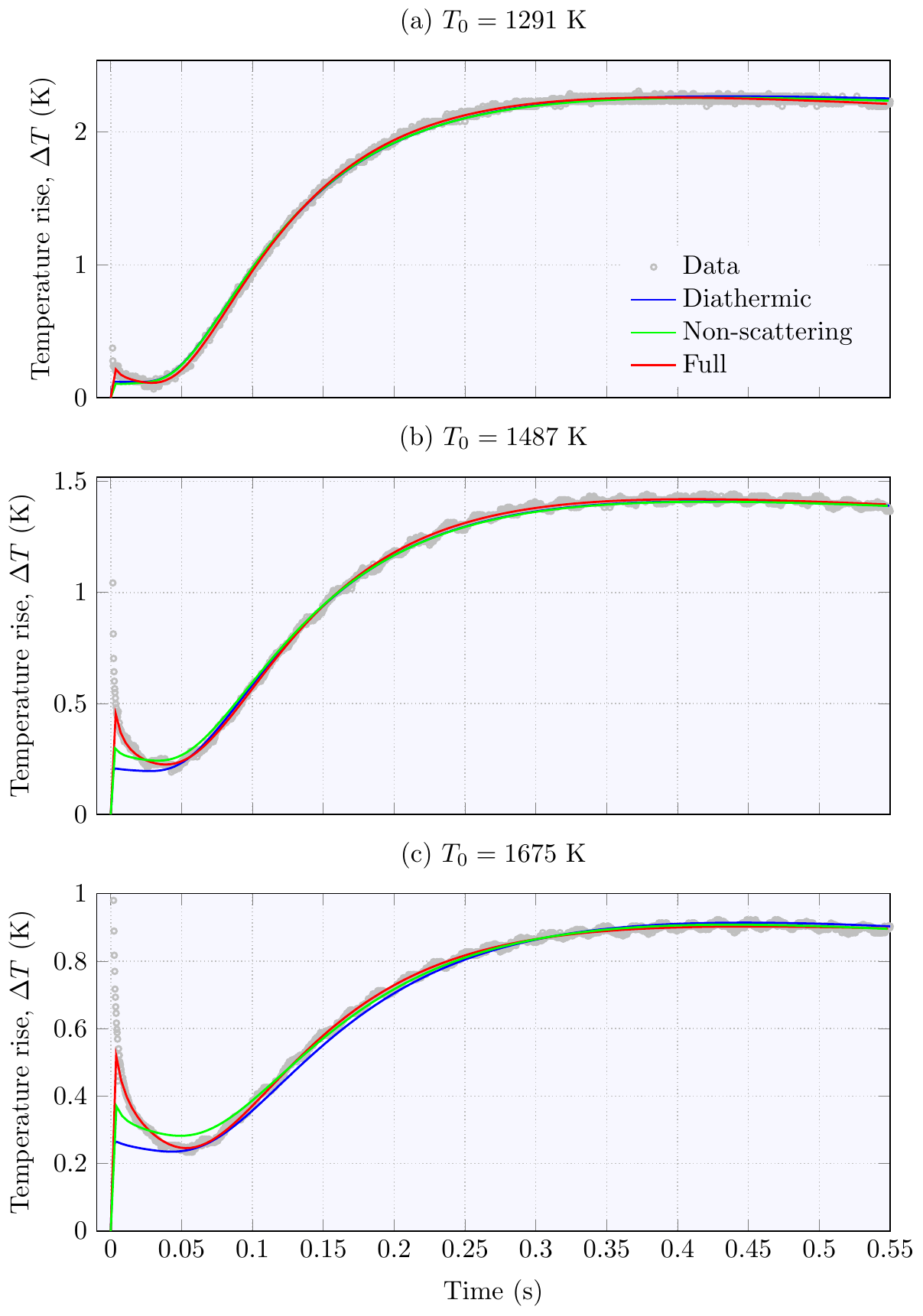}
	\caption{\label{fig:al2o3_high_temp} Experimental rear-surface time-temperature profiles of an Al$_2$O$_3$ sample initially thermalised at three different temperatures. Shown are the optimal solutions of the simplified diathermic model and of the fully-coupled radiative-conductive model.} 
\end{figure}

Finally, a set of experimental data acquired for a synthetic alumina sample~($l = 1.181$~mm) measured in a laser flash apparatus at high temperatures has been provided for validating the computational procedure. Measurements were conducted using the Kvant instrument at the Moscow Engineering Physics Institute, previously briefly described in~\cite{Lunev2020}. Specific heat and thermal expansion data have been taken from~\cite{ditmars1982enthalpy,Engberg1959}. Density at room temperature was measured using the hydrostatic method. Example time-temperature profiles are shown in~\cref{fig:al2o3_high_temp} along with the solutions to the inverse problem using three different models. A sharp temperature peak at the start of experiment is especially pronounced at the highest ambient temperatures. Only the complete calculation with the Henyey-Greenstein phase function is capable of reproducing this behaviour, although the model deviates from the experiment slightly at the start. Possibly this is due to some residual coating on the side surface of the sample which may have created an easy path for thermal diffusion. Another point to be aware of is the fact that the sample holder used in these experiments covered a significant area of the sample. The holder effectively consisted of two washers pressed against both sides of the sample while typically a three-point contact scheme is used in modern instruments. Thus, laser radiation was non-uniformly absorbed at the front surface, covering approximately~$75$~\%.   

\begin{figure}
\centering
\includegraphics{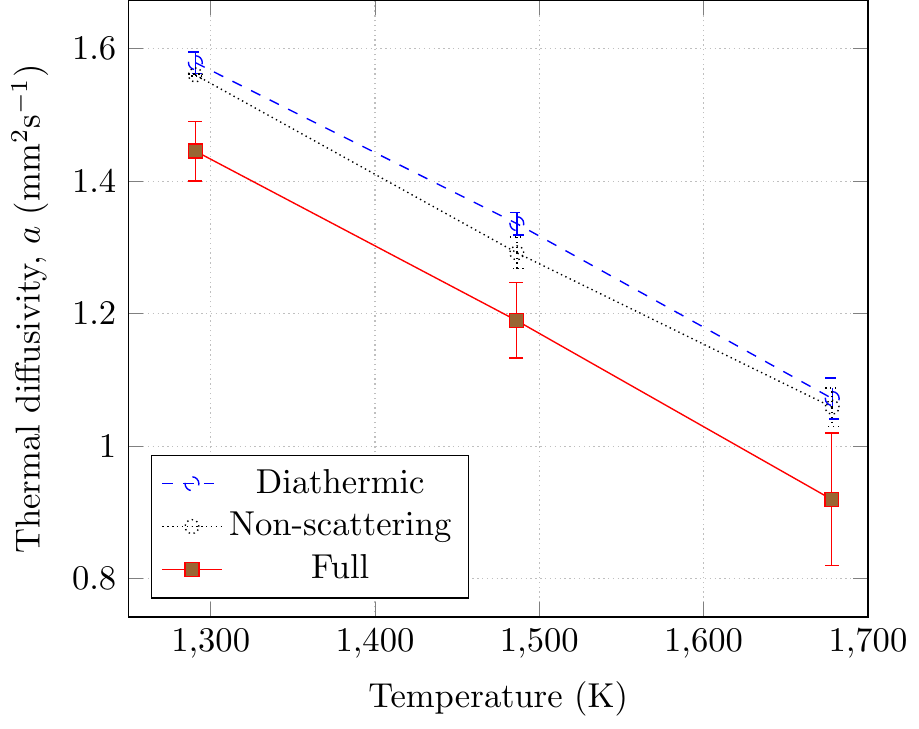}
\caption{\label{fig:results} Thermal diffusivity of synthetic alumina determined from a single set of experimental data using three different models. Note the uncertainties associated with the different starting parameters for the full model.}
\end{figure}

At each test temperature, thermal diffusivity~(\cref{fig:results}) was averaged over three measurements. Results show that a complete calculation produces systematically different values compared to the diathermic model with a maximum deviation of over~$10\%$. The high error margins are due to the optimisation procedure finding different minima depending on the starting conditions. The tendency of the optimiser to slip into a local minimum is due to the objective function being acute and multi-modal, which commonly occurs in multi-variate optimisation; moreover, even though the set of parameters can be sufficiently different, the minima are not. This highlights the necessity of introducing additional constraints -- relying on e.g. the optical properties.

\section{Conclusions}

The numerical method described in this paper combines: \begin{enumerate*}[label=(\alph*)]
	\item a stiffness-aware solver, its error control scheme and an adaptive stretching grid -- specifically tailored to solving the initial value problems arising from the discretised radiative transfer equation;
	\item a composite Gaussian quadrature designed to treat discontinuous intensities typical to the one-dimensional radiative transfer and 
	\item a fourth-order semi-implicit finite-difference scheme for numerically solving the heat problem
\end{enumerate*}. This combination is applied to enhance the data analysis in laser flash experiments where the material under study scatters thermal radiation anisotropically, such as when conducting measurements on transparent alumina at high temperatures. The calculation procedure reproduces the initial rapid variation of temperature typical to the strongly-scattering medium while still observing physically-reasonable values of secondary model parameters~(i.e., of the optical thickness, Planck number, emissivity, scattering albedo and of the anisotropic factor). The estimate quality is benchmarked against a standard diathermic model, where the maximum deviation is observed at high temperatures and pronounced scattering anisotropy. The optimisation procedure has been modified to implement constrained search using a one-to-one mapping of the search variables. This allowed imposing realistic parameter constraints. A further refinement of the search procedure is recommended to correctly address the ill-posed problems often occurring in multi-variate optimisation. The algorithms have been successfully implemented in the PULsE software, with the latest version being immediately available for use.

\section*{Acknowledgements}

\noindent This work was partially funded by the RCUK Energy Programme~(Grant No. EP/T012250/1). A. L. gratefully acknowledges the experimental dataset kindly provided by Dr. A. Tenishev~(MEPhI) and the involvement of Ms. A. Elbakyan in literature review.

\appendix
\section{Justification of using linearised boundary conditions}
\label{sect:appendix_a}

\begin{figure}
	\includegraphics{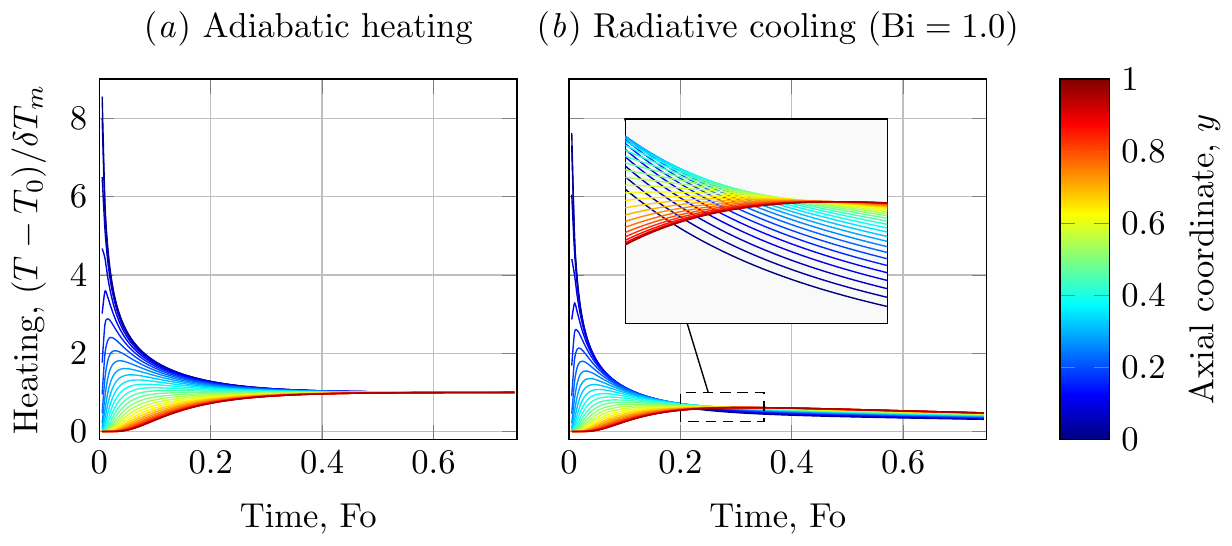}
	\caption{\label{fig:full_profile_1}Full time-temperature profile as calculated from the linearized one-dimensional problem after a~$\mathrm{Fo}_{\mathrm{las}} = 1 \times 10^{-4}$ pulse~(fully implicit scheme, $N=30$). Note the crossover of thermograms at different~$y$ for the radiative cooling.}
\end{figure}

For the sake of simplicity, the analysis is based on the same heat conduction problem as previously described in~\cite{Lunev2020}. An example distribution of the time-temperature profiles across the spatial domain is shown in \cref{fig:full_profile_1}. Clearly, the dimensionless temperature~$\theta$ can reach quite high values close to the front boundary~($y =0$), thus indicating a possible source of error in the conventional analysis, which assumes small heating~($T - T_0 \ll T_0$). The goal is to quantify that error. 

Omitting the heat equation and the initial condition, which are exactly the same as in~\cref{sect:diathermic}, the problem at hand is reduced to the following set of equations:

	\begin{subequations}
	\label{eq:boundary_problem_1d}
	\begin{align}
	\label{eq:he_1d_dim}
	&{\left. {\frac{{\partial T}}{{\partial z}}} \right|_{z = 0}} =  - \frac{4Q}{{\pi \lambda {d^2}}}P(t) + \frac{\varepsilon ({T_0})\sigma_0 T_0^4}{\lambda} \left \{ \left[ \frac{ {T_{z=0} - T_0} }{T_0} + 1 \right]^4 - 1 \right \},\\
	\label{eq:bc_lx_dim}
	&{\left. {\frac{{\partial T}}{{\partial z}}} \right|_{z = l}} =  - \frac{\varepsilon ({T_0})\sigma_0 T_0^4}{\lambda} \left \{ \left[ \frac{ {T_{z=l} - T_0} }{T_0} + 1 \right]^4 - 1 \right \},
	\end{align}	
\end{subequations}
where~$Q$ is the heat absorbed by the thin surface layer and~$\varepsilon$ is the sample's flat surface emissivity. These equations are then transformed to the dimensionless form:

\begin{subequations}
\label{eq:boundary_problem_1d_t}
\begin{align}
\label{eq:he_1d_dim_t}
&{\left. {\frac{{\partial \theta}}{{\partial y}}} \right|_{y = 0}} =  - \Phi(\mathrm{Fo}) + \mathrm{Bi} \cdot T_0 \left [ \left ( \theta_{y=0} \delta {T_{\mathrm{m}}} /{T_0} + 1 \right )^4 - 1 \right ]/( 4\delta {T_{\mathrm{m}}} ),\\
\label{eq:bc_lx_dim_t}
&{\left. {\frac{{\partial \theta}}{{\partial y}}} \right|_{y = 1}} =  - \mathrm{Bi} \cdot T_0 \left [ \left ( \theta_{y=1} \delta T_{\mathrm{m}} /{T_0} + 1 \right )^4 - 1 \right ]/(4 \delta {T_{\mathrm{m}}}),
\end{align}	
\end{subequations}
where~$\delta T_{\mathrm{m}}  = {4Q} ( \pi d^2 C_{\mathrm{p}} \rho l)^{-1}$ is the maximum heating of the rear surface in the absence of heat sinks and~${\rm{Bi}} := {{4\sigma_0 \varepsilon T_0^3l}}/{\lambda }$ is the Biot number, and~$\theta = (T - T_0)/\delta {T_{\mathrm{m}}}$. 

It can be easily seen that if~$\theta \delta T_{\mathrm{m}}/T_0$ is small, the heat loss term becomes simply~$\mathrm{Bi} \cdot \theta_y$, which corresponds to the classical case. When~$\delta T_{\mathrm{m}}/T_0 \simeq 1$, using only the first term of the Taylor expansion might not be appropriate; especially at the front surface~($y = 0$, see Fig.~\ref{fig:full_profile_1}), since~$\theta_{y=0} \gg \theta_{y=1}$ at~$\mathrm{Fo} = 0 - 0.15$. However, the overall magnitude of the heat sink term is proportional to~$T_0/4\delta T_m$. Hence, the significance of this term may be low when the expression in the brackets may be nonlinear.
	
The finite-difference calculations proceed as follows. The domain is divided into a uniform grid by introducing the coordinate step size~$h=1/(N-1)$, where~$N$ is the number of individual coordinate points on the grid, and the discrete time step~$\tau = \tau_{\mathrm{F}} h^2$, $\tau_{\mathrm{F}} \in \mathbb{R}$. The grid is used to discretise~$\theta(y,\rm{Fo})$, which becomes~$\theta(\xi_j,\widehat{\rm{Fo}}_m) = \theta_j^{m}$, $j = 0, ..., N-1$, $m = 0,...,m_0$, called the grid function. Let~$L \phi(\xi_{\alpha}) = \left ( \phi_{\alpha+1} - \phi_{\alpha-1} \right ) / 2h$. Then, the finite-difference analog of Eqs.~\ref{eq:boundary_problem_1d_t} is:

\begin{subequations}
	\label{eq:finite-difference}
	\begin{gather}
	\label{eq:fd_0}
	L \theta_0 = - \Xi + \zeta(\theta_0),\\
	\label{eq:fd_1}
	L \theta_{N-1} =  - \zeta(\theta_{N-1}), \\
	\zeta (\theta_j) = \mathrm{Bi} \cdot T_0/(4 \delta T_{\mathrm{m}}) \cdot \left [ (\theta_j \cdot \delta T_{\mathrm{m}}/T_0 + 1)^4 - 1 \right ],
	\end{gather}
\end{subequations}	
where the time index is implicit.	
	
Consider using a Taylor expansion on the grid at~$j=0$ and~$j=N-1$ and introducing virtual nodes~$j=-1$ and~$j = N$, thus transforming \cref{eq:finite-difference} using contraction mapping:~$\phi = \zeta(\phi)$. For a fully-implicit scheme the first coefficients~$\alpha_1$ and~$\beta_1$ from the tridiagonal matrix equation~$\theta_j = \alpha_{j+1} + \theta_{j+1} \beta_{j+1}$ and the solution at the~$j = N$ boundary are calculated at each iteration~$k + 1$ until the scheme converges to a given precision~(usually within a few iterations): 

\begin{subequations}
\begin{align}
&\stackrel{k+1}{[\alpha_1]} = \frac{2 \tau}{2 \tau + h^2}, \\
&\stackrel{k+1}{[\beta_1]} = \frac{h^2}{2\tau + h^2} \widehat{\theta}_0 + \frac{2 \tau h}{2 \tau + h^2} \left [ \Xi - \zeta(\stackrel{k}{\theta_0}) \right ],\\
&\stackrel{k+1}{[\theta_{N-1}]} = \frac{2 \tau \stackrel{k}{[\beta_{N-1}]} + h^2 \widehat{\theta}_{N-1} - 2 \tau h \zeta(\stackrel{k}{\theta_{N-1}})}{2 \tau + h^2 - 2 \tau \stackrel{k}{[\alpha_{N-1}]}},
\end{align}
\end{subequations}

\begin{figure}
	\centering
	\includegraphics{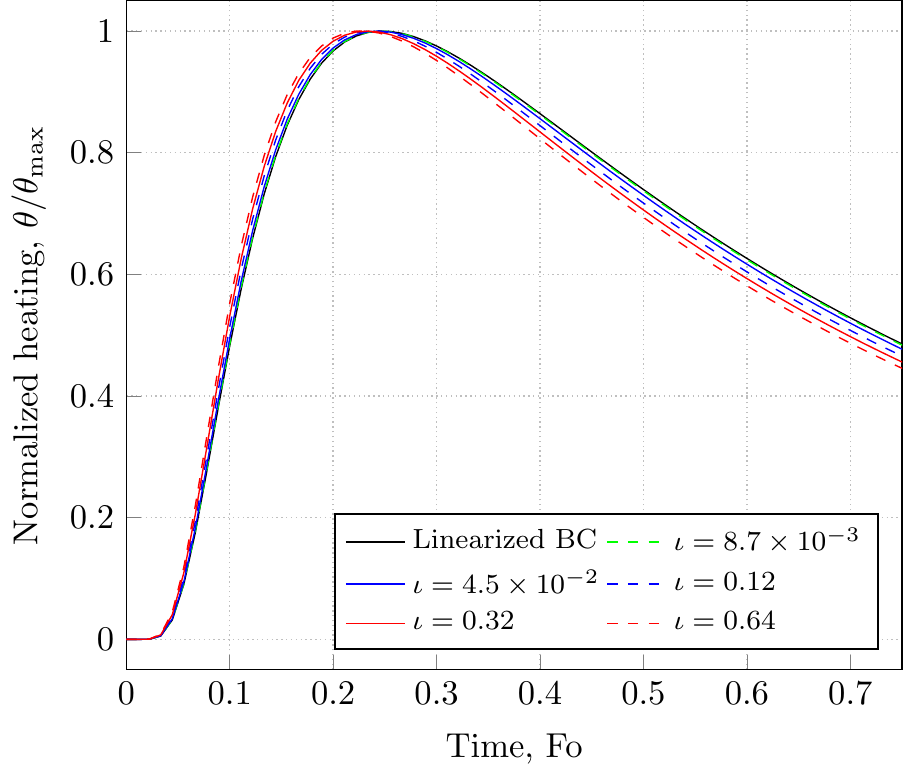}
	\caption{\label{fig:nonlinear_comparison} The effect of nonlinear heat losses~(Eqs.~(\ref{eq:boundary_problem_1d_t})) on the shape of the rear-surface heating curve evaluated by solving the boundary problem at different values of~$\iota = \delta T_{\mathrm{m}}/T_0$ using a fully-implicit finite-difference scheme and a fixed-point iteration algorithm~($\mathrm{Bi = 1.0}$, $\mathrm{Fo}_{\mathrm{las}} \approx 5 \times 10^{-3}$, fixed-point error tolerance~$\Delta_1 = 10^{-8}$~K).}	
\end{figure}

The solution is shown in~\cref{fig:nonlinear_comparison} where the heating curves have been normalized. Curves are plotted at different values of~$\iota: = \delta T_{\mathrm{max}}/T_0$, all else being equal. With increasing the~$r$ factor, the normalized maximum shifts towards shorter times while the temperature decreases due to heat losses~(in this case, $\mathrm{Bi} = 1.0$) becomes more pronounced. For~$\delta T_{\mathrm{max}}/T_0 < 5 \times 10^{-2}$~(in most practical cases), this effect is so small that the nonlinear behaviour of the heat losses in \cref{eq:boundary_problem_1d_t} may be completely neglected. Therefore, some care must be taken only when conducting measurements at cryogenic temperatures and at a high laser power applied to poor thermal conductors. Otherwise, keeping nonlinear terms in the boundary conditions is redundant and a simpler~(linearised) model of the heat problem may be used instead.

%
%
%
%
%
%
%
\section{Numerical evaluation of some integrals}
\label{sect:appendix_b}

The integrand function~$E_1(t)$ is discontinuous at~$t=0$, which complicates the evaluation of radiative flux derivatives~$dq/d\tau$ using the standard Newton-Cotes formulae. The latter require significant computational resources, which is inappropriate when the flux derivatives need to be calculated frequently. 

The general problem consists in evaluating integrals of the form:

\begin{equation}
\label{eq:complex_integral}
I_n = \int_a^b {g(t) E_n(\alpha + \beta t) dt}.
\end{equation}

The exponential integrals~$E_n(t)$ are pre-calculated using the midpoint rule with a very large number of integration points by filling a look-up table of typically~$N_{\mathrm{tab}} = 10,000 - 20,000$ entries, depending on the cutoff value~($t_c^{\mathrm{exp}} = 9.2 - 21.0$), which ensures a precision of at least~$10^{-5}$. This table is filled only once at the program start and used later in future calls to the solver. An acceptable accuracy when using a Newton-Cotes formula (e.g. the Simpson's rule) can be achieved at~$n_q=256$~[see~\cref{tbl:quadratures}] for integrals of order~$n \geq 2$ when the integrand is well-defined at zero. Since the exponential integrals rapidly decrease with~$\tau$ and the emission function~$j(t)$ is bounded, the integrand becomes very small where the exponential integrals are near-zero. The integration bounds are calculated as~$[\max\{a, (t_c - \alpha)/\beta\}, b]$ at~$\beta < 0$ and~$[a, \min\{b, (t_c - \alpha)/\beta\}]$ at~$\beta > 0$. This ensures that for large~$\tau_0$, the integration excludes terms smaller in amplitude than a certain threshold defined by the cutoff~$t_c$. Additionally, since~$f(t)$ is discretised differently to what is used in the quadrature scheme, a natural cubic spline interpolation implemented in the The Apache Commons Mathematics Library is introduced to calculate the function values.  

A more effective quadrature has been introduced by~\citet{Chandrasekhar1960}. It is first noticed that~\cref{eq:complex_integral} may be written as:

\begin{align}
\label{eq:chandrasekhar}
\int_{\alpha + \beta a}^{\alpha + \beta b} { g\left ( \beta (x - \alpha) \right ) E_n(x) dx} = \sum_{j = 1}^{m}{a_j g(x_j)}.
\end{align}. 

The moments~$M_l$ are defined as:

\begin{align}
\label{eq:moments}
M_l = \int_{\alpha + \beta a}^{\alpha + \beta b} { x^l E_n(x) dx }.
\end{align} 

These can be integrated by parts if the recurrent expression for~$E_n(x)$ is utilised~\cite{Chandrasekhar1960}. After the moments have been calculated, the next step is to calculate the~$x_j$~($j = 1,..,m$) roots of the monic polynomial $x^m + \sum_{l = 0}^{m-1}{c_l x^l}$ where the coefficients~$c_l$ form the solution of a linear set:

\begin{equation}
M_{i+m} + \sum_{l=0}^{m-1}{c_l M_{i+l}} = 0, \quad i = 0,1,...,m-1.
\end{equation}

In fact, the latter is effectively a matrix equation, which may simply be solved using matrix inversion. The roots~$x_j$ are then found with the help of a Laguerre solver implemented in the Apache Commons Mathematics Library. The weights~$a_j$ of the quadrature~\cref{eq:chandrasekhar} should satisfy the~$m$ equations:

\begin{align}
M_l = \sum_{j = 1}^m{a_j x_j^l}, \quad l = 0,...,m-1.
\end{align} 

This is solved in a similar fashion. \Cref{tbl:quadratures,tbl:quadratures_i1} show test results of using the Chandrasekhar's quadrature versus the Newton-Cotes formulae. These test have been carried out for a test temperature profile shown in~\cref{fig:test_profile}.   

\begin{table}
	\centering
	\caption{Comparison between quadrature formulae for calculating $I_2 = \int_0^{\tau_0}{j[\theta(t)]E_2(\alpha + \beta t) dt}$ at~$\tau_0 = 2.0$, $\beta = -1$, $\alpha = \tau_0$ using a test temperature profile.}
	
	\label{tbl:quadratures}
	\begin{tabular}{rrclrrc}
		\hline
		\multicolumn{3}{c}{Simpson's rule}                                                                                              &  & \multicolumn{3}{c}{Chandrasekhar's quadrature}                                                                                  \\
		\multicolumn{1}{c}{$n$} & \multicolumn{1}{c}{$I_2$} & $\Delta$                     &  & \multicolumn{1}{c}{$m$} & \multicolumn{1}{c}{$I_2$} & $\Delta$                     \\ \cline{1-3} \cline{5-7} 
		32                      & 940.70148                                                              & -                            &  & 2                       & 940.10042                                                              & -                            \\
		256                     & 940.10960                                                              & $-0.59190$  &  & 4                       & 940.09943                                                              & $-0.00112$ \\
		4096                    & 940.10074                                                              & $-0.00886$ &  & 8                       & 940.09948                                                              & $+0.00005$  \\ \hline
	\end{tabular}%
\end{table}

\begin{table}
	\centering
	\caption{Comparison of end precision~$\Delta$ and computational effort~$T_{10,000}$~(measured for $10,000$ consecutive calls to the respective integration method) for different quadrature formulae for calculating the integral $I_1 = \int_{\tau}^{\tau_0}{j[\theta(t)]E_1(\alpha + \beta t) dt}$ at~$\tau_0 = 3.0$, $\beta = 1$, $\tau = -\alpha = 0.5$ using the same test temperature profile as in~\Cref{tbl:quadratures}.}
	\label{tbl:quadratures_i1}
	\resizebox{\textwidth}{!}{%
		\begin{tabular}{rrcrlrrcr}
			\hline
			\multicolumn{4}{c}{Simpson's rule}                                                                                                                                    &                      & \multicolumn{4}{c}{Chandrasekhar's quadrature}                                                                             \\
			\multicolumn{1}{c}{$n$} & \multicolumn{1}{c}{$I_1$} & $\Delta$                     & \multicolumn{1}{l}{\begin{tabular}[c]{@{}l@{}}$T_{10,000}$\\  (ms)\end{tabular}} & \multicolumn{1}{c}{} & \multicolumn{1}{c}{$m$} & \multicolumn{1}{c}{$I_1$} & $\Delta$                     & \multicolumn{1}{l}{$T_{10,000}$ (ms)} \\ \cline{1-4} \cline{6-9} 
			32                      & 2190.51                   & -                            & 20                                                                               &                      & 2                       & 1961.618                  & -                            & 82                                    \\
			256                     & 1976.71                   & \multicolumn{1}{l}{$-213.8$} & 121                                                                              &                      & 3                       & 1961.617                  & \multicolumn{1}{r}{$-0.001$} & 163                                   \\
			4096                    & 1962.31                   & \multicolumn{1}{r}{$-14.4$}  & 1254                                                                             &                      & 8                       & 1961.617                  & \multicolumn{1}{r}{0}        & 620                                   \\ \hline
		\end{tabular}%
	}
\end{table}

	\newpage
	
\bibliography{Bibliography}

\begin{thebibliography}{81}
\expandafter\ifx\csname natexlab\endcsname\relax\def\natexlab#1{#1}\fi
\providecommand{\url}[1]{\texttt{#1}}
\providecommand{\href}[2]{#2}
\providecommand{\path}[1]{#1}
\providecommand{\DOIprefix}{doi:}
\providecommand{\ArXivprefix}{arXiv:}
\providecommand{\URLprefix}{URL: }
\providecommand{\Pubmedprefix}{pmid:}
\providecommand{\doi}[1]{\href{http://dx.doi.org/#1}{\path{#1}}}
\providecommand{\Pubmed}[1]{\href{pmid:#1}{\path{#1}}}
\providecommand{\bibinfo}[2]{#2}
\ifx\xfnm\relax \def\xfnm[#1]{\unskip,\space#1}\fi
\bibitem[{Pavlov et~al.(2017)Pavlov, Wenman, Vlahovic, Robba, Konings, {V}an
  {U}ffelen, and Grimes}]{Pavlov2017}
\bibinfo{author}{T.~Pavlov}, \bibinfo{author}{M.~Wenman},
  \bibinfo{author}{L.~Vlahovic}, \bibinfo{author}{D.~Robba},
  \bibinfo{author}{R.~Konings}, \bibinfo{author}{P.~{V}an {U}ffelen},
  \bibinfo{author}{R.~Grimes},
\newblock \bibinfo{title}{Measurement and interpretation of the thermo-physical
  properties of {UO}$_2$ at high temperatures: The viral effect of oxygen
  defects},
\newblock \bibinfo{journal}{Acta Materialia} \bibinfo{volume}{139}
  (\bibinfo{year}{2017}) \bibinfo{pages}{138 -- 154}.
  \DOIprefix\doi{https://doi.org/10.1016/j.actamat.2017.07.060}.
\bibitem[{Zhao et~al.(2019)Zhao, Sun, Li, Xie, Meng, Wang, and
  Zhang}]{Zhao2019}
\bibinfo{author}{S.~Zhao}, \bibinfo{author}{X.~Sun}, \bibinfo{author}{Z.~Li},
  \bibinfo{author}{W.~Xie}, \bibinfo{author}{S.~Meng},
  \bibinfo{author}{C.~Wang}, \bibinfo{author}{W.~Zhang},
\newblock \bibinfo{title}{Simultaneous retrieval of high temperature thermal
  conductivities, anisotropic radiative properties, and thermal contact
  resistance for ceramic foams},
\newblock \bibinfo{journal}{Applied Thermal Engineering} \bibinfo{volume}{146}
  (\bibinfo{year}{2019}) \bibinfo{pages}{569 -- 576}.
  \DOIprefix\doi{https://doi.org/10.1016/j.applthermaleng.2018.10.021}.
\bibitem[{Lunev and Heymer(2020)}]{Lunev2020}
\bibinfo{author}{A.~Lunev}, \bibinfo{author}{R.~Heymer},
\newblock \bibinfo{title}{Decreasing the uncertainty of classical laser flash
  analysis using numerical algorithms robust to noise and systematic errors},
\newblock \bibinfo{journal}{Review of Scientific Instruments}
  \bibinfo{volume}{91} (\bibinfo{year}{2020}) \bibinfo{pages}{064902}.
  \DOIprefix\doi{10.1063/1.5132786}.
\bibitem[{Olorunyolemi et~al.(2002)Olorunyolemi, Birnboim, Carmel, Wilson~Jr.,
  Lloyd, Smith, and Campbell}]{Olo2002}
\bibinfo{author}{T.~Olorunyolemi}, \bibinfo{author}{A.~Birnboim},
  \bibinfo{author}{Y.~Carmel}, \bibinfo{author}{O.~C. Wilson~Jr.},
  \bibinfo{author}{I.~K. Lloyd}, \bibinfo{author}{S.~Smith},
  \bibinfo{author}{R.~Campbell},
\newblock \bibinfo{title}{Thermal conductivity of zinc oxide: From green to
  sintered state},
\newblock \bibinfo{journal}{Journal of the American Ceramic Society}
  \bibinfo{volume}{85} (\bibinfo{year}{2002}) \bibinfo{pages}{1249--1253}.
  \DOIprefix\doi{10.1111/j.1151-2916.2002.tb00253.x}.
\bibitem[{Itatani et~al.(2006)Itatani, Tsujimoto, and Kishimoto}]{Ita2006}
\bibinfo{author}{K.~Itatani}, \bibinfo{author}{T.~Tsujimoto},
  \bibinfo{author}{A.~Kishimoto},
\newblock \bibinfo{title}{Thermal and optical properties of transparent
  magnesium oxide ceramics fabricated by post hot-isostatic pressing},
\newblock \bibinfo{journal}{Journal of the European Ceramic Society}
  \bibinfo{volume}{26} (\bibinfo{year}{2006}) \bibinfo{pages}{639 -- 645}.
  \DOIprefix\doi{https://doi.org/10.1016/j.jeurceramsoc.2005.06.011},
  \bibinfo{note}{proceedings of the International Symposium on Inorganic and
  Environmental Materials, Eindhoven, NL, October 2004}.
\bibitem[{Cozzo et~al.(2011)Cozzo, Staicu, Somers, Fernandez, and
  Konings}]{Cozzo2011}
\bibinfo{author}{C.~Cozzo}, \bibinfo{author}{D.~Staicu},
  \bibinfo{author}{J.~Somers}, \bibinfo{author}{A.~Fernandez},
  \bibinfo{author}{R.~Konings},
\newblock \bibinfo{title}{Thermal diffusivity and conductivity of
  thorium–plutonium mixed oxides},
\newblock \bibinfo{journal}{Journal of Nuclear Materials} \bibinfo{volume}{416}
  (\bibinfo{year}{2011}) \bibinfo{pages}{135 -- 141}.
  \DOIprefix\doi{https://doi.org/10.1016/j.jnucmat.2011.01.109},
  \bibinfo{note}{nuclear Materials IV}.
\bibitem[{Bison et~al.(2007)Bison, Cernuschi, Grinzato, Marinetti, and
  Robba}]{Bison2007}
\bibinfo{author}{P.~Bison}, \bibinfo{author}{F.~Cernuschi},
  \bibinfo{author}{E.~Grinzato}, \bibinfo{author}{S.~Marinetti},
  \bibinfo{author}{D.~Robba},
\newblock \bibinfo{title}{Ageing evaluation of thermal barrier coatings by
  thermal diffusivity},
\newblock \bibinfo{journal}{Infrared Physics \& Technology}
  \bibinfo{volume}{49} (\bibinfo{year}{2007}) \bibinfo{pages}{286 -- 291}.
  \DOIprefix\doi{https://doi.org/10.1016/j.infrared.2006.06.019}.
\bibitem[{Tischler et~al.(1988)Tischler, Kohanoff, Rangugni, and
  Ondracek}]{Tischler1988}
\bibinfo{author}{M.~Tischler}, \bibinfo{author}{J.~J. Kohanoff},
  \bibinfo{author}{G.~A. Rangugni}, \bibinfo{author}{G.~Ondracek},
\newblock \bibinfo{title}{Pulse method of measuring thermal diffusivity and
  optical absorption depth for partially transparent materials},
\newblock \bibinfo{journal}{Journal of Applied Physics} \bibinfo{volume}{63}
  (\bibinfo{year}{1988}) \bibinfo{pages}{1259--1264}.
  \DOIprefix\doi{10.1063/1.339950}.
\bibitem[{McMasters et~al.(1999)McMasters, Beck, Dinwiddie, and
  Wang}]{McMasters1999}
\bibinfo{author}{R.~L. McMasters}, \bibinfo{author}{J.~V. Beck},
  \bibinfo{author}{R.~B. Dinwiddie}, \bibinfo{author}{H.~Wang},
\newblock \bibinfo{title}{{Accounting for Penetration of Laser Heating in Flash
  Thermal Diffusivity Experiments}},
\newblock \bibinfo{journal}{Journal of Heat Transfer} \bibinfo{volume}{121}
  (\bibinfo{year}{1999}) \bibinfo{pages}{15--21}.
  \DOIprefix\doi{10.1115/1.2825929}.
\bibitem[{Blumm et~al.(1997)Blumm, Henderson, Nilsson, and
  Fricke}]{blumm1997laser}
\bibinfo{author}{J.~Blumm}, \bibinfo{author}{J.~B. Henderson},
  \bibinfo{author}{O.~Nilsson}, \bibinfo{author}{J.~Fricke},
\newblock \bibinfo{title}{Laser flash measurement of the phononic thermal
  diffusivity of glasses in the presence of ballistic radiative transfer},
\newblock \bibinfo{journal}{High Temperatures. High Pressures (Print)}
  \bibinfo{volume}{29} (\bibinfo{year}{1997}) \bibinfo{pages}{555--560}.
\bibitem[{Mehling et~al.(1998)Mehling, Hautzinger, Nilsson, Fricke, Hofmann,
  and Hahn}]{Mehling1998}
\bibinfo{author}{H.~Mehling}, \bibinfo{author}{G.~Hautzinger},
  \bibinfo{author}{O.~Nilsson}, \bibinfo{author}{J.~Fricke},
  \bibinfo{author}{R.~Hofmann}, \bibinfo{author}{O.~Hahn},
\newblock \bibinfo{title}{Thermal diffusivity of semitransparent materials
  determined by the laser-flash method applying a new analytical model},
\newblock \bibinfo{journal}{International Journal of Thermophysics}
  \bibinfo{volume}{19} (\bibinfo{year}{1998}) \bibinfo{pages}{941--949}.
\bibitem[{Andre and Degiovanni(1995)}]{Andre1995}
\bibinfo{author}{S.~Andre}, \bibinfo{author}{A.~Degiovanni},
\newblock \bibinfo{title}{A theoretical study of the transient coupled
  conduction and radiation heat transfer in glass: phonic diffusivity
  measurements by the flash technique},
\newblock \bibinfo{journal}{International Journal of Heat and Mass Transfer}
  \bibinfo{volume}{38} (\bibinfo{year}{1995}) \bibinfo{pages}{3401 -- 3412}.
  \DOIprefix\doi{https://doi.org/10.1016/0017-9310(95)00075-K}.
\bibitem[{Andre and Degiovanni(1998)}]{Andre1998}
\bibinfo{author}{S.~Andre}, \bibinfo{author}{A.~Degiovanni},
\newblock \bibinfo{title}{{A New Way of Solving Transient Radiative-Conductive
  Heat Transfer Problems}},
\newblock \bibinfo{journal}{Journal of Heat Transfer} \bibinfo{volume}{120}
  (\bibinfo{year}{1998}) \bibinfo{pages}{943--955}.
  \DOIprefix\doi{10.1115/1.2825914}.
\bibitem[{Lazard et~al.(2001{\natexlab{a}})Lazard, Andre, Maillet, Baillis, and
  Degiovanni}]{Lazard2001}
\bibinfo{author}{M.~Lazard}, \bibinfo{author}{S.~Andre},
  \bibinfo{author}{D.~Maillet}, \bibinfo{author}{D.~Baillis},
  \bibinfo{author}{A.~Degiovanni},
\newblock \bibinfo{title}{Flash experiment on a semitransparent material:
  interest of a reduced model},
\newblock \bibinfo{journal}{Inverse Problems in Engineering}
  \bibinfo{volume}{9} (\bibinfo{year}{2001}{\natexlab{a}})
  \bibinfo{pages}{413--429}. \DOIprefix\doi{10.1080/174159701088027772}.
\bibitem[{Lazard et~al.(2001{\natexlab{b}})Lazard, André, and
  Maillet}]{Lazard2001_2}
\bibinfo{author}{M.~Lazard}, \bibinfo{author}{S.~André},
  \bibinfo{author}{D.~Maillet},
\newblock \bibinfo{title}{Transient coupled radiative–conductive heat
  transfer in a gray planar medium with anisotropic scattering},
\newblock \bibinfo{journal}{Journal of Quantitative Spectroscopy and Radiative
  Transfer} \bibinfo{volume}{69} (\bibinfo{year}{2001}{\natexlab{b}})
  \bibinfo{pages}{23 -- 33}.
  \DOIprefix\doi{https://doi.org/10.1016/S0022-4073(00)00054-6}.
\bibitem[{Braiek et~al.(2016)Braiek, Adili, Albouchi, Karkri, and
  Nasrallah}]{Braiek2016}
\bibinfo{author}{A.~Braiek}, \bibinfo{author}{A.~Adili},
  \bibinfo{author}{F.~Albouchi}, \bibinfo{author}{M.~Karkri},
  \bibinfo{author}{S.~B. Nasrallah},
\newblock \bibinfo{title}{Estimation of radiative and conductive properties of
  a semitransparent medium using genetic algorithms},
\newblock \bibinfo{journal}{Measurement Science and Technology}
  \bibinfo{volume}{27} (\bibinfo{year}{2016}) \bibinfo{pages}{065601}.
  \DOIprefix\doi{10.1088/0957-0233/27/6/065601}.
\bibitem[{Modest and Azad(1980)}]{Modest1980}
\bibinfo{author}{M.~F. Modest}, \bibinfo{author}{F.~H. Azad},
\newblock \bibinfo{title}{{The Influence and Treatment of Mie-Anisotropic
  Scattering in Radiative Heat Transfer}},
\newblock \bibinfo{journal}{Journal of Heat Transfer} \bibinfo{volume}{102}
  (\bibinfo{year}{1980}) \bibinfo{pages}{92--98}.
  \DOIprefix\doi{10.1115/1.3244255}.
\bibitem[{Hahn et~al.(1997)Hahn, Raether, Arduini-Schuster, and
  Fricke}]{Hahn1997}
\bibinfo{author}{O.~Hahn}, \bibinfo{author}{F.~Raether},
  \bibinfo{author}{M.~Arduini-Schuster}, \bibinfo{author}{J.~Fricke},
\newblock \bibinfo{title}{Transient coupled conductive/radiative heat transfer
  in absorbing, emitting and scattering media: application to laser-flash
  measurements on ceramic materials},
\newblock \bibinfo{journal}{International Journal of Heat and Mass Transfer}
  \bibinfo{volume}{40} (\bibinfo{year}{1997}) \bibinfo{pages}{689 -- 698}.
  \DOIprefix\doi{https://doi.org/10.1016/0017-9310(96)00137-8}.
\bibitem[{Chandrasekhar(1960)}]{Chandrasekhar1960}
\bibinfo{author}{S.~Chandrasekhar}, \bibinfo{title}{Radiative transfer},
  \bibinfo{publisher}{Dover Publications (New York, NY)}, \bibinfo{year}{1960}.
\bibitem[{da~Silva et~al.(1998)da~Silva, Laurent, and
  Baillis-Doermann}]{Silva1998}
\bibinfo{author}{Z.~da~Silva}, \bibinfo{author}{M.~Laurent},
  \bibinfo{author}{D.~Baillis-Doermann},
\newblock \bibinfo{title}{Inverse analysis of transient coupled
  conduction-radiation-conductive and radiative properties and measurements},
\newblock in: \bibinfo{booktitle}{7th AIAA/ASME Joint Thermophysics and Heat
  Transfer Conference}, \bibinfo{year}{1998}, p. \bibinfo{pages}{2842}.
\bibitem[{Coquard et~al.(2009)Coquard, Rochais, and Baillis}]{Coquard2009}
\bibinfo{author}{R.~Coquard}, \bibinfo{author}{D.~Rochais},
  \bibinfo{author}{D.~Baillis},
\newblock \bibinfo{title}{Experimental investigations of the coupled conductive
  and radiative heat transfer in metallic/ceramic foams},
\newblock \bibinfo{journal}{International Journal of Heat and Mass Transfer}
  \bibinfo{volume}{52} (\bibinfo{year}{2009}) \bibinfo{pages}{4907 -- 4918}.
  \DOIprefix\doi{https://doi.org/10.1016/j.ijheatmasstransfer.2009.05.015}.
\bibitem[{Coquard et~al.(2011)Coquard, Randrianalisoa, Lallich, and
  Baillis}]{Coquard2011}
\bibinfo{author}{R.~Coquard}, \bibinfo{author}{J.~Randrianalisoa},
  \bibinfo{author}{S.~Lallich}, \bibinfo{author}{D.~Baillis},
\newblock \bibinfo{title}{{Extension of the FLASH Method to Semitransparent
  Polymer Foams}},
\newblock \bibinfo{journal}{Journal of Heat Transfer} \bibinfo{volume}{133}
  (\bibinfo{year}{2011}). \DOIprefix\doi{10.1115/1.4004392}.
\bibitem[{Wellele et~al.(2006)Wellele, Orlande, Ruperti, Colaço, and
  Delmas}]{Wel2006}
\bibinfo{author}{O.~Wellele}, \bibinfo{author}{H.~Orlande},
  \bibinfo{author}{N.~Ruperti}, \bibinfo{author}{M.~Colaço},
  \bibinfo{author}{A.~Delmas},
\newblock \bibinfo{title}{Coupled conduction–radiation in semi-transparent
  materials at high temperatures},
\newblock \bibinfo{journal}{Journal of Physics and Chemistry of Solids}
  \bibinfo{volume}{67} (\bibinfo{year}{2006}) \bibinfo{pages}{2230 -- 2240}.
  \DOIprefix\doi{https://doi.org/10.1016/j.jpcs.2006.06.007},
  \bibinfo{note}{sMEC 2005}.
\bibitem[{Sans et~al.(2020)Sans, Schick, Parent, and Farges}]{Sans2020}
\bibinfo{author}{M.~Sans}, \bibinfo{author}{V.~Schick},
  \bibinfo{author}{G.~Parent}, \bibinfo{author}{O.~Farges},
\newblock \bibinfo{title}{Experimental characterization of the coupled
  conductive and radiative heat transfer in ceramic foams with a flash method
  at high temperature},
\newblock \bibinfo{journal}{International Journal of Heat and Mass Transfer}
  \bibinfo{volume}{148} (\bibinfo{year}{2020}) \bibinfo{pages}{119077}.
  \DOIprefix\doi{https://doi.org/10.1016/j.ijheatmasstransfer.2019.119077}.
\bibitem[{Zmywaczyk and Koniorczyk(2009)}]{zmywaczyk2009numerical}
\bibinfo{author}{J.~Zmywaczyk}, \bibinfo{author}{P.~Koniorczyk},
\newblock \bibinfo{title}{Numerical solution of inverse radiative--conductive
  transient heat transfer problem in a grey participating medium},
\newblock \bibinfo{journal}{International Journal of Thermophysics}
  \bibinfo{volume}{30} (\bibinfo{year}{2009}) \bibinfo{pages}{1438--1451}.
\bibitem[{Lacroix et~al.(2002)Lacroix, Parent, Asllanaj, and
  Jeandel}]{Lacrois2002}
\bibinfo{author}{D.~Lacroix}, \bibinfo{author}{G.~Parent},
  \bibinfo{author}{F.~Asllanaj}, \bibinfo{author}{G.~Jeandel},
\newblock \bibinfo{title}{Coupled radiative and conductive heat transfer in a
  non-grey absorbing and emitting semitransparent media under collimated
  radiation},
\newblock \bibinfo{journal}{Journal of Quantitative Spectroscopy and Radiative
  Transfer} \bibinfo{volume}{75} (\bibinfo{year}{2002}) \bibinfo{pages}{589 --
  609}. \DOIprefix\doi{https://doi.org/10.1016/S0022-4073(02)00031-6}.
\bibitem[{Fiveland(1984)}]{Fiveland1984}
\bibinfo{author}{W.~A. Fiveland},
\newblock \bibinfo{title}{{Discrete-Ordinates Solutions of the Radiative
  Transport Equation for Rectangular Enclosures}},
\newblock \bibinfo{journal}{Journal of Heat Transfer} \bibinfo{volume}{106}
  (\bibinfo{year}{1984}) \bibinfo{pages}{699--706}.
  \DOIprefix\doi{10.1115/1.3246741}.
\bibitem[{Fiveland(1987)}]{Fiveland1987}
\bibinfo{author}{W.~A. Fiveland},
\newblock \bibinfo{title}{{Discrete Ordinate Methods for Radiative Heat
  Transfer in Isotropically and Anisotropically Scattering Media}},
\newblock \bibinfo{journal}{Journal of Heat Transfer} \bibinfo{volume}{109}
  (\bibinfo{year}{1987}) \bibinfo{pages}{809--812}.
  \DOIprefix\doi{10.1115/1.3248167}.
\bibitem[{Philipp et~al.(2020)Philipp, Eichinger, Aydin, Georgiadis, Cyron, and
  Retsch}]{philipp2020accuracy}
\bibinfo{author}{A.~Philipp}, \bibinfo{author}{J.~F. Eichinger},
  \bibinfo{author}{R.~C. Aydin}, \bibinfo{author}{A.~Georgiadis},
  \bibinfo{author}{C.~J. Cyron}, \bibinfo{author}{M.~Retsch},
\newblock \bibinfo{title}{The accuracy of laser flash analysis explored by
  finite element method and numerical fitting},
\newblock \bibinfo{journal}{Heat and Mass Transfer} \bibinfo{volume}{56}
  (\bibinfo{year}{2020}) \bibinfo{pages}{811--823}.
\bibitem[{Coelho(2008)}]{Coelho2007}
\bibinfo{author}{P.~Coelho},
\newblock \bibinfo{title}{A comparison of spatial discretization schemes for
  differential solution methods of the radiative transfer equation},
\newblock \bibinfo{journal}{Journal of Quantitative Spectroscopy and Radiative
  Transfer} \bibinfo{volume}{109} (\bibinfo{year}{2008}) \bibinfo{pages}{189 --
  200}. \DOIprefix\doi{https://doi.org/10.1016/j.jqsrt.2007.08.012},
  \bibinfo{note}{the Fifth International Symposium on Radiative Transfer}.
\bibitem[{Coelho(2014)}]{Coelho2014}
\bibinfo{author}{P.~J. Coelho},
\newblock \bibinfo{title}{Advances in the discrete ordinates and finite volume
  methods for the solution of radiative heat transfer problems in participating
  media},
\newblock \bibinfo{journal}{Journal of Quantitative Spectroscopy and Radiative
  Transfer} \bibinfo{volume}{145} (\bibinfo{year}{2014}) \bibinfo{pages}{121 --
  146}. \DOIprefix\doi{https://doi.org/10.1016/j.jqsrt.2014.04.021}.
\bibitem[{Lunev(2020)}]{software}
\bibinfo{author}{A.~Lunev}, \bibinfo{title}{kotik-coder/{PULsE}: {PULsE}
  v1.79}, \bibinfo{year}{2020}. \URLprefix
  \url{https://doi.org/10.5281/zenodo.3928762}.
  \DOIprefix\doi{10.5281/zenodo.3928762}.
\bibitem[{Yoffa(1980)}]{Yoffa1980}
\bibinfo{author}{E.~J. Yoffa},
\newblock \bibinfo{title}{Role of carrier diffusion in lattice heating during
  pulsed laser annealing},
\newblock \bibinfo{journal}{Applied Physics Letters} \bibinfo{volume}{36}
  (\bibinfo{year}{1980}) \bibinfo{pages}{37--38}.
  \DOIprefix\doi{10.1063/1.91306}.
\bibitem[{Howell et~al.(2010)Howell, Menguc, and Siegel}]{howell2010thermal}
\bibinfo{author}{J.~R. Howell}, \bibinfo{author}{M.~P. Menguc},
  \bibinfo{author}{R.~Siegel}, \bibinfo{title}{Thermal radiation heat
  transfer}, \bibinfo{publisher}{CRC press}, \bibinfo{year}{2010}.
\bibitem[{Samarskii and Nikolaev(1978)}]{samarskii1978methods}
\bibinfo{author}{A.~Samarskii}, \bibinfo{author}{E.~Nikolaev},
  \bibinfo{title}{Methods of solving finite-difference equations},
  \bibinfo{publisher}{Nauka, Moscow}, \bibinfo{year}{1978}.
\bibitem[{Kourganoff(1963)}]{kourganoff1963basic}
\bibinfo{author}{V.~Kourganoff}, \bibinfo{title}{Basic Methods in Transfer
  Problems}, \bibinfo{publisher}{Dover}, \bibinfo{year}{1963}.
\bibitem[{Modest(2013)}]{modest2013radiative}
\bibinfo{author}{M.~F. Modest}, \bibinfo{title}{Radiative heat transfer},
  \bibinfo{publisher}{Academic press}, \bibinfo{year}{2013}.
\bibitem[{Cess(1964)}]{Cess1964}
\bibinfo{author}{R.~Cess},
\newblock \bibinfo{title}{The interaction of thermal radiation with conduction
  and convection heat transfer},
\newblock in: \bibinfo{editor}{T.~F. Irvine}, \bibinfo{editor}{J.~P. Hartnett}
  (Eds.), \bibinfo{booktitle}{Advances in Heat Transfer},
  volume~\bibinfo{volume}{1}, \bibinfo{publisher}{Elsevier},
  \bibinfo{year}{1964}, pp. \bibinfo{pages}{1 -- 50}.
  \DOIprefix\doi{https://doi.org/10.1016/S0065-2717(08)70096-0}.
\bibitem[{Schuster(1905)}]{schuster1905radiation}
\bibinfo{author}{A.~Schuster},
\newblock \bibinfo{title}{Radiation through a foggy atmosphere},
\newblock \bibinfo{journal}{The astrophysical journal} \bibinfo{volume}{21}
  (\bibinfo{year}{1905}) \bibinfo{pages}{1}.
\bibitem[{Schwartzschild and Gesell(1906)}]{schwartzschild1906gottingen}
\bibinfo{author}{K.~Schwartzschild}, \bibinfo{author}{W.~Gesell},
\newblock \bibinfo{title}{{G}ottingen, {N}achr. {M}ath},
\newblock \bibinfo{journal}{Phys. Klasse}  (\bibinfo{year}{1906})
  \bibinfo{pages}{41}.
\bibitem[{Brewster and Tien(1982)}]{Brewster1982}
\bibinfo{author}{M.~Brewster}, \bibinfo{author}{C.~Tien},
\newblock \bibinfo{title}{Examination of the two-flux model for radiative
  transfer in particular systems},
\newblock \bibinfo{journal}{International Journal of Heat and Mass Transfer}
  \bibinfo{volume}{25} (\bibinfo{year}{1982}) \bibinfo{pages}{1905 -- 1907}.
  \DOIprefix\doi{https://doi.org/10.1016/0017-9310(82)90113-2}.
\bibitem[{Mengüç and Viskanta(1983)}]{Menguc1983}
\bibinfo{author}{M.~Mengüç}, \bibinfo{author}{R.~Viskanta},
\newblock \bibinfo{title}{Comparison of radiative transfer approximations for a
  highly forward scattering planar medium},
\newblock \bibinfo{journal}{Journal of Quantitative Spectroscopy and Radiative
  Transfer} \bibinfo{volume}{29} (\bibinfo{year}{1983}) \bibinfo{pages}{381 --
  394}. \DOIprefix\doi{https://doi.org/10.1016/0022-4073(83)90111-5}.
\bibitem[{Engler(2015)}]{Engler2015}
\bibinfo{author}{H.~Engler},
\newblock \bibinfo{title}{Computation of scattering kernels in radiative
  transfer},
\newblock \bibinfo{journal}{Journal of Quantitative Spectroscopy and Radiative
  Transfer} \bibinfo{volume}{165} (\bibinfo{year}{2015}) \bibinfo{pages}{38 --
  42}. \DOIprefix\doi{https://doi.org/10.1016/j.jqsrt.2015.06.019}.
\bibitem[{van~de Hulst(1981)}]{hulst1981light}
\bibinfo{author}{H.~C. van~de Hulst}, \bibinfo{title}{Light scattering by small
  particles}, \bibinfo{publisher}{Courier Corporation}, \bibinfo{year}{1981}.
\bibitem[{Henyey and Greenstein(1941)}]{henyey1941diffuse}
\bibinfo{author}{L.~G. Henyey}, \bibinfo{author}{J.~L. Greenstein},
\newblock \bibinfo{title}{Diffuse radiation in the galaxy},
\newblock \bibinfo{journal}{The {A}strophysical {J}ournal} \bibinfo{volume}{93}
  (\bibinfo{year}{1941}) \bibinfo{pages}{70--83}.
\bibitem[{Kattawar(1975)}]{kattawar1975three}
\bibinfo{author}{G.~W. Kattawar},
\newblock \bibinfo{title}{A three-parameter analytic phase function for
  multiple scattering calculations},
\newblock \bibinfo{journal}{Journal of Quantitative Spectroscopy and Radiative
  Transfer} \bibinfo{volume}{15} (\bibinfo{year}{1975})
  \bibinfo{pages}{839--849}.
\bibitem[{Haltrin(2002)}]{Haltrin}
\bibinfo{author}{V.~I. Haltrin},
\newblock \bibinfo{title}{One-parameter two-term {H}enyey-{G}reenstein phase
  function for light scattering in seawater},
\newblock \bibinfo{journal}{Applied Optics} \bibinfo{volume}{41}
  (\bibinfo{year}{2002}) \bibinfo{pages}{1022--1028}.
\bibitem[{Wang et~al.(2019)Wang, Xu, Nilsson, Fernandes, and
  Niklasson}]{Wang2019}
\bibinfo{author}{J.~Wang}, \bibinfo{author}{C.~Xu}, \bibinfo{author}{A.~M.
  Nilsson}, \bibinfo{author}{D.~L.~A. Fernandes}, \bibinfo{author}{G.~A.
  Niklasson},
\newblock \bibinfo{title}{A novel phase function describing light scattering of
  layers containing colloidal nanospheres},
\newblock \bibinfo{journal}{Nanoscale} \bibinfo{volume}{11}
  (\bibinfo{year}{2019}) \bibinfo{pages}{7404--7413}.
  \DOIprefix\doi{10.1039/C9NR01707K}.
\bibitem[{Zhao et~al.(2019)Zhao, Sun, Que, and Zhang}]{zhao2019influence}
\bibinfo{author}{S.~Zhao}, \bibinfo{author}{X.~Sun}, \bibinfo{author}{Q.~Que},
  \bibinfo{author}{W.~Zhang},
\newblock \bibinfo{title}{Influence of scattering phase function on estimated
  thermal properties of {Al}$_2${O}$_3$ ceramic foams},
\newblock \bibinfo{journal}{International Journal of Thermophysics}
  \bibinfo{volume}{40} (\bibinfo{year}{2019}) \bibinfo{pages}{11}.
\bibitem[{Boulet et~al.(2007)Boulet, Collin, and Consalvi}]{Boulet2007}
\bibinfo{author}{P.~Boulet}, \bibinfo{author}{A.~Collin},
  \bibinfo{author}{J.~Consalvi},
\newblock \bibinfo{title}{On the finite volume method and the discrete
  ordinates method regarding radiative heat transfer in acute forward
  anisotropic scattering media},
\newblock \bibinfo{journal}{Journal of Quantitative Spectroscopy and Radiative
  Transfer} \bibinfo{volume}{104} (\bibinfo{year}{2007}) \bibinfo{pages}{460 --
  473}. \DOIprefix\doi{https://doi.org/10.1016/j.jqsrt.2006.09.010}.
\bibitem[{Samarskii(2001)}]{samarskii2001theory}
\bibinfo{author}{A.~A. Samarskii}, \bibinfo{title}{The theory of difference
  schemes}, volume \bibinfo{volume}{240}, \bibinfo{publisher}{CRC Press},
  \bibinfo{year}{2001}.
\bibitem[{Jessee and Fiveland(1997)}]{Jessee1997}
\bibinfo{author}{J.~P. Jessee}, \bibinfo{author}{W.~A. Fiveland},
\newblock \bibinfo{title}{Bounded, high-resolution differencing schemes applied
  to the discrete ordinates method},
\newblock \bibinfo{journal}{Journal of Thermophysics and Heat Transfer}
  \bibinfo{volume}{11} (\bibinfo{year}{1997}) \bibinfo{pages}{540--548}.
  \DOIprefix\doi{10.2514/2.6296}.
\bibitem[{Liu et~al.(1996)Liu, Becker, and Pollard}]{Liu1996}
\bibinfo{author}{F.~Liu}, \bibinfo{author}{H.~A. Becker},
  \bibinfo{author}{A.~Pollard},
\newblock \bibinfo{title}{Spatial differencing schemes of the
  discrete-ordinates method},
\newblock \bibinfo{journal}{Numerical Heat Transfer, Part B: Fundamentals}
  \bibinfo{volume}{30} (\bibinfo{year}{1996}) \bibinfo{pages}{23--43}.
  \DOIprefix\doi{10.1080/10407799608915070}.
\bibitem[{Maginot et~al.(2016)Maginot, Ragusa, and Morel}]{Maginot2016}
\bibinfo{author}{P.~G. Maginot}, \bibinfo{author}{J.~C. Ragusa},
  \bibinfo{author}{J.~E. Morel},
\newblock \bibinfo{title}{High-order solution methods for grey discrete
  ordinates thermal radiative transfer},
\newblock \bibinfo{journal}{Journal of Computational Physics}
  \bibinfo{volume}{327} (\bibinfo{year}{2016}) \bibinfo{pages}{719 -- 746}.
  \DOIprefix\doi{https://doi.org/10.1016/j.jcp.2016.09.055}.
\bibitem[{Alexander(1977)}]{Alexander1977}
\bibinfo{author}{R.~Alexander},
\newblock \bibinfo{title}{Diagonally implicit {R}unge–{K}utta methods for
  stiff {O.D.E.}’s},
\newblock \bibinfo{journal}{SIAM Journal on Numerical Analysis}
  \bibinfo{volume}{14} (\bibinfo{year}{1977}) \bibinfo{pages}{1006--1021}.
  \DOIprefix\doi{10.1137/0714068}.
\bibitem[{D’Alessandro et~al.(2018)D’Alessandro, Binci, Montelpare, and
  Ricci}]{Alessandro2018}
\bibinfo{author}{V.~D’Alessandro}, \bibinfo{author}{L.~Binci},
  \bibinfo{author}{S.~Montelpare}, \bibinfo{author}{R.~Ricci},
\newblock \bibinfo{title}{On the development of openfoam solvers based on
  explicit and implicit high-order {R}unge–{K}utta schemes for incompressible
  flows with heat transfer},
\newblock \bibinfo{journal}{Computer Physics Communications}
  \bibinfo{volume}{222} (\bibinfo{year}{2018}) \bibinfo{pages}{14 -- 30}.
  \DOIprefix\doi{https://doi.org/10.1016/j.cpc.2017.09.009}.
\bibitem[{Boom and Zingg(2018)}]{Boom2018}
\bibinfo{author}{P.~D. Boom}, \bibinfo{author}{D.~W. Zingg},
\newblock \bibinfo{title}{Optimization of high-order diagonally-implicit
  {R}unge–{K}utta methods},
\newblock \bibinfo{journal}{Journal of Computational Physics}
  \bibinfo{volume}{371} (\bibinfo{year}{2018}) \bibinfo{pages}{168 -- 191}.
  \DOIprefix\doi{https://doi.org/10.1016/j.jcp.2018.05.020}.
\bibitem[{Hairer and Wanner(1996)}]{hairer1996stiff}
\bibinfo{author}{E.~Hairer}, \bibinfo{author}{G.~Wanner},
  \bibinfo{title}{Solving Ordinary Differential Equations II. Stiff and
  Differential-Algebraic Problems}, volume~\bibinfo{volume}{14},
  \bibinfo{year}{1996}. \DOIprefix\doi{10.1007/978-3-662-09947-6}.
\bibitem[{Kennedy and Carpenter(2016)}]{kennedy2016diagonally}
\bibinfo{author}{C.~A. Kennedy}, \bibinfo{author}{M.~H. Carpenter},
  \bibinfo{title}{Diagonally implicit {R}unge-{K}utta methods for ordinary
  differential equations. A review}, \bibinfo{type}{Technical Report}, NASA
  Langley Research Center, Hampton, VA, United States, \bibinfo{year}{2016}.
\bibitem[{Blom et~al.(2016)Blom, Birken, Bijl, Kessels, Meister, and van
  Zuijlen}]{blom2016comparison}
\bibinfo{author}{D.~S. Blom}, \bibinfo{author}{P.~Birken},
  \bibinfo{author}{H.~Bijl}, \bibinfo{author}{F.~Kessels},
  \bibinfo{author}{A.~Meister}, \bibinfo{author}{A.~H. van Zuijlen},
\newblock \bibinfo{title}{A comparison of {R}osenbrock and {ESDIRK} methods
  combined with iterative solvers for unsteady compressible flows},
\newblock \bibinfo{journal}{Advances in Computational Mathematics}
  \bibinfo{volume}{42} (\bibinfo{year}{2016}) \bibinfo{pages}{1401--1426}.
\bibitem[{Dormand and Prince(1980)}]{DormanPrince54}
\bibinfo{author}{J.~Dormand}, \bibinfo{author}{P.~Prince},
\newblock \bibinfo{title}{A family of embedded {R}unge-{K}utta formulae},
\newblock \bibinfo{journal}{Journal of Computational and Applied Mathematics}
  \bibinfo{volume}{6} (\bibinfo{year}{1980}) \bibinfo{pages}{19 -- 26}.
  \DOIprefix\doi{https://doi.org/10.1016/0771-050X(80)90013-3}.
\bibitem[{Bogacki and Shampine(1989)}]{BS32}
\bibinfo{author}{P.~Bogacki}, \bibinfo{author}{L.~F. Shampine},
\newblock \bibinfo{title}{A 3 (2) pair of {R}unge-{K}utta formulas},
\newblock \bibinfo{journal}{Applied Mathematics Letters} \bibinfo{volume}{2}
  (\bibinfo{year}{1989}) \bibinfo{pages}{321--325}.
\bibitem[{Hairer et~al.(1993)Hairer, N{\o}rsett, and
  Wanner}]{hairer1993solving}
\bibinfo{author}{E.~Hairer}, \bibinfo{author}{S.~P. N{\o}rsett},
  \bibinfo{author}{G.~Wanner}, \bibinfo{title}{Solving ordinary differential
  equations I. {N}onstiff problems}, \bibinfo{publisher}{Springer Series in
  Computational Mathematics}, \bibinfo{year}{1993}.
\bibitem[{Demmel(1997)}]{demmel1997applied}
\bibinfo{author}{J.~W. Demmel}, \bibinfo{title}{Applied numerical linear
  algebra}, volume~\bibinfo{volume}{56}, \bibinfo{publisher}{Siam},
  \bibinfo{year}{1997}.
\bibitem[{Hosea and Shampine(1996)}]{TRBDF2}
\bibinfo{author}{M.~Hosea}, \bibinfo{author}{L.~Shampine},
\newblock \bibinfo{title}{Analysis and implementation of {TR}-{BDF}2},
\newblock \bibinfo{journal}{Applied Numerical Mathematics} \bibinfo{volume}{20}
  (\bibinfo{year}{1996}) \bibinfo{pages}{21 -- 37}.
  \DOIprefix\doi{https://doi.org/10.1016/0168-9274(95)00115-8},
  \bibinfo{note}{method of Lines for Time-Dependent Problems}.
\bibitem[{Nitsche(1996)}]{Stretching1}
\bibinfo{author}{L.~C. Nitsche},
\newblock \bibinfo{title}{One-dimensional stretching functions for cn patched
  grids, and associated truncation errors in finite-difference calculations},
\newblock \bibinfo{journal}{Communications in Numerical Methods in Engineering}
  \bibinfo{volume}{12} (\bibinfo{year}{1996}) \bibinfo{pages}{303--316}.
  \DOIprefix\doi{10.1002/(SICI)1099-0887(199605)12:5<303::AID-CNM979>3.0.CO;2-C}.
\bibitem[{Vinokur(1983)}]{Stretching2}
\bibinfo{author}{M.~Vinokur},
\newblock \bibinfo{title}{On one-dimensional stretching functions for
  finite-difference calculations},
\newblock \bibinfo{journal}{Journal of Computational Physics}
  \bibinfo{volume}{50} (\bibinfo{year}{1983}) \bibinfo{pages}{215 -- 234}.
  \DOIprefix\doi{https://doi.org/10.1016/0021-9991(83)90065-7}.
\bibitem[{Rogers and Adams(1989)}]{rogers1989mathematical}
\bibinfo{author}{D.~F. Rogers}, \bibinfo{author}{J.~A. Adams},
  \bibinfo{title}{Mathematical elements for computer graphics},
  \bibinfo{publisher}{McGraw-Hill Higher Education}, \bibinfo{year}{1989}.
\bibitem[{Lathrop and Carlson(1964)}]{lathrop1964discrete}
\bibinfo{author}{K.~D. Lathrop}, \bibinfo{author}{B.~G. Carlson},
  \bibinfo{title}{Discrete ordinates angular quadrature of the neutron
  transport equation}, \bibinfo{type}{Technical Report}, Los Alamos Scientific
  Lab., N. Mex., \bibinfo{year}{1964}.
\bibitem[{Truelove(1987)}]{Truelove1987}
\bibinfo{author}{J.~S. Truelove},
\newblock \bibinfo{title}{{Discrete-Ordinate Solutions of the Radiation
  Transport Equation}},
\newblock \bibinfo{journal}{Journal of Heat Transfer} \bibinfo{volume}{109}
  (\bibinfo{year}{1987}) \bibinfo{pages}{1048--1051}.
  \DOIprefix\doi{10.1115/1.3248182}.
\bibitem[{Kumar et~al.(1990)Kumar, Majumdar, and Tien}]{Kumar1990}
\bibinfo{author}{S.~Kumar}, \bibinfo{author}{A.~Majumdar},
  \bibinfo{author}{C.~L. Tien},
\newblock \bibinfo{title}{{The Differential-Discrete-Ordinate Method for
  Solutions of the Equation of Radiative Transfer}},
\newblock \bibinfo{journal}{Journal of Heat Transfer} \bibinfo{volume}{112}
  (\bibinfo{year}{1990}) \bibinfo{pages}{424--429}.
  \DOIprefix\doi{10.1115/1.2910395}.
\bibitem[{Li et~al.(1998)Li, Yao, Cao, and Cen}]{Li1998}
\bibinfo{author}{B.-W. Li}, \bibinfo{author}{Q.~Yao}, \bibinfo{author}{X.-Y.
  Cao}, \bibinfo{author}{K.-F. Cen},
\newblock \bibinfo{title}{{A New Discrete Ordinates Quadrature Scheme for
  Three-Dimensional Radiative Heat Transfer}},
\newblock \bibinfo{journal}{Journal of Heat Transfer} \bibinfo{volume}{120}
  (\bibinfo{year}{1998}) \bibinfo{pages}{514--518}.
  \DOIprefix\doi{10.1115/1.2824279}.
\bibitem[{Liu et~al.(2002)Liu, Ruan, and Tan}]{Liu2002}
\bibinfo{author}{L.~Liu}, \bibinfo{author}{L.~Ruan}, \bibinfo{author}{H.~Tan},
\newblock \bibinfo{title}{On the discrete ordinates method for radiative heat
  transfer in anisotropically scattering media},
\newblock \bibinfo{journal}{International Journal of Heat and Mass Transfer}
  \bibinfo{volume}{45} (\bibinfo{year}{2002}) \bibinfo{pages}{3259 -- 3262}.
  \DOIprefix\doi{https://doi.org/10.1016/S0017-9310(02)00035-2}.
\bibitem[{Koch and Becker(2004)}]{Koch2004}
\bibinfo{author}{R.~Koch}, \bibinfo{author}{R.~Becker},
\newblock \bibinfo{title}{Evaluation of quadrature schemes for the discrete
  ordinates method},
\newblock \bibinfo{journal}{Journal of Quantitative Spectroscopy and Radiative
  Transfer} \bibinfo{volume}{84} (\bibinfo{year}{2004}) \bibinfo{pages}{423 --
  435}. \DOIprefix\doi{https://doi.org/10.1016/S0022-4073(03)00260-7},
  \bibinfo{note}{{E}urotherm {S}eminar 73 - {C}omputational {T}hermal
  {R}adiation in {P}articipating {M}edia}.
\bibitem[{Lebedev(1976)}]{Lebedev1976}
\bibinfo{author}{V.~Lebedev},
\newblock \bibinfo{title}{Quadratures on a sphere},
\newblock \bibinfo{journal}{USSR Computational Mathematics and Mathematical
  Physics} \bibinfo{volume}{16} (\bibinfo{year}{1976}) \bibinfo{pages}{10 --
  24}. \DOIprefix\doi{https://doi.org/10.1016/0041-5553(76)90100-2}.
\bibitem[{Thynell(1998)}]{Thynell1998}
\bibinfo{author}{S.~T. Thynell},
\newblock \bibinfo{title}{Discrete-ordinates method in radiative heat
  transfer},
\newblock \bibinfo{journal}{International Journal of Engineering Science}
  \bibinfo{volume}{36} (\bibinfo{year}{1998}) \bibinfo{pages}{1651 -- 1675}.
  \DOIprefix\doi{https://doi.org/10.1016/S0020-7225(98)00052-4}.
\bibitem[{Dombrovsky et~al.(2011)Dombrovsky, Randrianalisoa, Lipiński, and
  Baillis}]{Fresnel2011}
\bibinfo{author}{L.~A. Dombrovsky}, \bibinfo{author}{J.~H. Randrianalisoa},
  \bibinfo{author}{W.~Lipiński}, \bibinfo{author}{D.~Baillis},
\newblock \bibinfo{title}{Approximate analytical solution to normal emittance
  of semi-transparent layer of an absorbing, scattering, and refracting
  medium},
\newblock \bibinfo{journal}{Journal of Quantitative Spectroscopy and Radiative
  Transfer} \bibinfo{volume}{112} (\bibinfo{year}{2011}) \bibinfo{pages}{1987
  -- 1994}. \DOIprefix\doi{https://doi.org/10.1016/j.jqsrt.2011.04.008}.
\bibitem[{Tikhonov(1963)}]{tikhonov1963solution}
\bibinfo{author}{A.~N. Tikhonov},
\newblock \bibinfo{title}{On the solution of ill-posed problems and the method
  of regularization},
\newblock in: \bibinfo{booktitle}{Doklady Akademii Nauk}, volume
  \bibinfo{volume}{151}, \bibinfo{organization}{Russian Academy of Sciences},
  \bibinfo{year}{1963}, pp. \bibinfo{pages}{501--504}.
\bibitem[{Gill et~al.(2019)Gill, Murray, and Wright}]{gill2019practical}
\bibinfo{author}{P.~E. Gill}, \bibinfo{author}{W.~Murray},
  \bibinfo{author}{M.~H. Wright}, \bibinfo{title}{Practical optimization},
  \bibinfo{publisher}{SIAM}, \bibinfo{year}{2019}.
\bibitem[{Ditmars et~al.(1982)Ditmars, Ishihara, Chang, Bernstein, and
  West}]{ditmars1982enthalpy}
\bibinfo{author}{D.~Ditmars}, \bibinfo{author}{S.~Ishihara},
  \bibinfo{author}{S.~Chang}, \bibinfo{author}{G.~Bernstein},
  \bibinfo{author}{E.~West},
\newblock \bibinfo{title}{Enthalpy and heat-capacity standard reference
  material: synthetic sapphire ($\alpha$-{A}l$_2${O}$_3$) from 10 to 2250 k},
\newblock \bibinfo{journal}{Journal of Research of the National Bureau of
  Standards} \bibinfo{volume}{87} (\bibinfo{year}{1982})
  \bibinfo{pages}{159--63}.
\bibitem[{Engberg and Zaehms(1959)}]{Engberg1959}
\bibinfo{author}{C.~J. Engberg}, \bibinfo{author}{E.~H. Zaehms},
\newblock \bibinfo{title}{Thermal expansion of {A}l$_2${O}$_3$, {B}e{O},
  {M}g{O}, {B}$_4${C}, {S}i{C}, and {T}i{C} above 1000{°}c.},
\newblock \bibinfo{journal}{Journal of the American Ceramic Society}
  \bibinfo{volume}{42} (\bibinfo{year}{1959}) \bibinfo{pages}{300--305}.
  \DOIprefix\doi{10.1111/j.1151-2916.1959.tb12958.x}.

\end{thebibliography}
	
\end{document}